\documentclass[twocolumn]{aastex62}

\usepackage{savesym} \savesymbol{tablenum} \usepackage{siunitx} \restoresymbol{SIX}{tablenum} 
\usepackage{amsmath}

\newcommand{\lowtran}{{\tt lowtran7}}
\newcommand{\python}{{\tt python}}
\newcommand{\scivision}{\href{https://github.com/scivision/lowtran}{this URL}}
\newcommand{\hirs}{\href{ftp.ssec.wisc.edu}{this FTP}}
\newcommand{\thisurl}{\href{http://spiff.rit.edu/classes/phys445/lectures/atmos/atmos.html}{this URL}}

\received{April 5, 2019}
\revised{July 12, 2019}
\accepted{July 18, 2019}
\submitjournal{PASP}

\shorttitle{The Terrascope}
\shortauthors{Kipping}

\begin{document}

\title{THE ``TERRASCOPE'': ON THE POSSIBILITY OF USING THE EARTH AS AN ATMOSPHERIC LENS}

\correspondingauthor{David Kipping}
\email{dkipping@astro.columbia.edu}

\author[0000-0002-4365-7366]{David Kipping}
\affil{Department of Astronomy, Columbia University, 550 W 120th Street,
New York, NY 10027, USA}



\begin{abstract}
Distant starlight passing through the Earth's atmosphere is refracted by an
angle of just over one degree near the surface.
This focuses light onto a focal line starting at an inner (and chromatic)
boundary out to infinity, offering an opportunity for pronounced lensing.
It is shown here that the focal line commences at ${\sim}85$\% of the
Earth-Moon separation, and thus placing an orbiting detector between here and
one Hill radius could exploit this refractive lens.
Analytic estimates are derived for a source directly behind the Earth (i.e.
on-axis) showing that starlight is lensed into a thin circular ring of
thickness $W H_{\Delta}/R$, yielding an amplification of
$8 H_{\Delta}/W$, where $H_{\Delta}$ is the Earth's refractive scale height,
$R$ is its geopotential radius, and $W$ is the detector diameter.
These estimates are verified through numerical ray-tracing experiments from
optical to \SI{30}{\micro\metre} light with standard atmospheric models.
The numerical experiments are extended to include extinction from both a
clear atmosphere and one with clouds. It is found that a detector at
one Hill radius is least affected by extinction since lensed rays travel
no deeper than 13.7\,km, within the statosphere and above most clouds.
Including extinction, a 1\,metre Hill radius ``terrascope'' is calculated to
produce an amplification of ${\sim} 45,000$ for a lensing timescale of ${\sim} 20$\,hours. In practice,
the amplification is likely halved in order to avoid daylight scattering
i.e. $22,500$ ($\Delta$mag=10.9) for $W=$\SI{1}{\metre}, or equivalent to a
\SI{150}{\metre} optical/infrared telescope.
\end{abstract}



\section{Introduction}
\label{sec:intro}

Astronomers crave photons. Simple Poisson counting statistics dictate that the
signal-to-noise (S/N) of any astronomical observation relying on electromagnetic
waves is proportional to the square root of the number of photons received per
unit time. Because the received photon rate is proportional to a telescope's
area, then S/N generally scales with telescope diameter.
Additionally, a larger telescope offers improved angular resolution, which is
is inversely proportional to the diameter.

These two benefits therefore both scale approximately linearly
with telescope diameter, yet the cost scales - at best - quadratically
\citep{belle:2004}. Besides the cost of the mirror itself, monolithic
mirrors larger than 5\,metres become difficult to build, due to large structures
needed to support them.
Segmented designs have therefore been proposed for our largest planned
telescopes, such as the \SI{25}{\metre} GMT, the \SI{30}{\metre} TMT and the
\SI{39}{\metre} ELT. With GMT priced at \$$1$B and the TMT at \$$2$B,
``post-1980'' scaling \citep{belle:2004} implies that a \SI{100}{\metre}
telescope would cost ${\sim}$\$$35$B, which greatly exceeds the combined 2018
budget of NASA and NSF. A similar situation is true for space-based
observatories, where the ${\sim}$\$$10$B \SI{6.5}{\metre} JWST provides an example of
how the funding prospects of yet larger telescopes look questionable - a point
described as a ``crisis in astrophysics'' \citep{elvis:2016}.

Faced with the prospect of a stall in our continuously improving view of the
Universe, it is timely to ask whether there exist any ``shortcuts'' to
catching photons - ways of amplifying the signal without building ever larger
structures. Angular resolution can certainly be improved using an
interferometric array of small telescopes rather than a single giant
structure \citep{monnier:2003}. Photon counts are much more challenging to
greatly increase, since the flux count per unit area is defined by an
astronomical source's magnitude, implying the only way to collect more photons
is to have a larger lenses. If we are unable to build such giant lenses
due to
their prohibitive cost, the next best thing is to ask if there exists any
naturally occurring giant lenses that could serve our purpose? 

One example of a natural lens is the Sun. Gravitational deflection and
lensing of light by massive objects was predicted by \citet{einstein:1916} and
observed for the Sun nearly a century ago \citep{eddington:1919}.
\citet{von:1979} proposed that this lensing could be used for interstellar
communication, with subsequent studies considering using the lens for astronomical
observations \citep{kraus:1986,heidmann:1994}.
Star light would be focussed along a line starting at 550\,AU and thus
an observatory would need to orbit the Sun beyond this separation, where
it would enjoy enormous amplification of distant sources. Besides launch operations,
there are challenges with such a mission.
It has been argued that interference from the solar atmospheric
plasma and limited source selection would pose major challenges to realizing
such a proposal \citep{turyshev:2003}.

When faced with the prospect of flying to 550\,AU to exploit an astrophysical
lens, it is natural to ask whether there are any alternative astrophysical
lenses closer to home? The Earth clearly cannot serve as a practical
gravitational lens given its relatively low mass, but it does bend light using
another physical effect - refraction. A setting Sun is a little more than half
a degree lower than it appears as a result of this effect. It therefore stands
to reason that a optical ray passing through the Earth's atmosphere, skimming
above the surface, and then emerging out the otherside will refract by just
over one degree. Thus, one might imagine an observer at or beyond a distance of
${\sim} R \cot$\SI{1}{\degree} $\simeq$ \SI{360000}{\kilo\metre}
(approximately the Earth-Moon separation) would be able to exploit the Earth as
a refractive lens - a concept referred to as the ``terrascope'' in what
follows. Much like the gravitational lens, this distance is the inner point of
a focal line, along which high amplification should be expected.

Refraction through the Earth's atmosphere to approximately one lunar separation
has been known since the $18^{\mathrm{th}}$ century \citep{cassini:1740},
through studies of the lunar eclipse. The use of this for lensing is first
hinted at by \citet{von:1979} who frame gravitational lensing as a refractive
analog. The concept is also apparent in images recorded by Apollo 12 crew
during an Earth-Sun eclipse, revealing lensing around the Earth's rim at large
separations \citep{apollo12:1969}. Atmospheric lensing is also utilized in
occultation measurements, in particular via the observable ``central flash''
such as that recorded for Neptune by \citet{hubbard:1987}. The idea of using
the Earth as a lens was next briefly discussed in \citet{wang:1998}, although
no estimates for the amplification were offered. This work aims to provide the
first quantitative grounding for the terrascope concept by calculating the
amplification expected, lens properties, image shapes, lensing timescale and
the effects of extinction. As an initial quantification of these effects, the
goal here is not to perform the most realistic calculation possible, but rather
estimate the approximate effects that might be expected. Other effects not
modeled, such as airglow, pointing stability and turbulence are briefly explored
in the discussion.

The paper is organized as follows. In Section~\ref{sec:refraction} a numerical
scheme for calculating refraction is described, as well as a simplified
atmospheric model. In Section~\ref{sec:raytracing}, ray tracing simulations are
described detailing how lensing solutions may be solved for. In
Section~\ref{sec:results}, the simulations are used to compute the
amplification expected for the terrascope as a function of wavelength,
observatory distance and detector diameter. This section also describes
estimates for extinction effects such as scattering and clouds.
Section~\ref{sec:discussion} concludes by describing ignored effects and
practical considerations.

\section{Modeling Atmospheric Refraction}
\label{sec:refraction}

\subsection{Lensing through a 1D static atmosphere}
\label{sub:refraction}

\begin{figure*}
\begin{center}
\includegraphics[width=17.0cm,angle=0,clip=true]{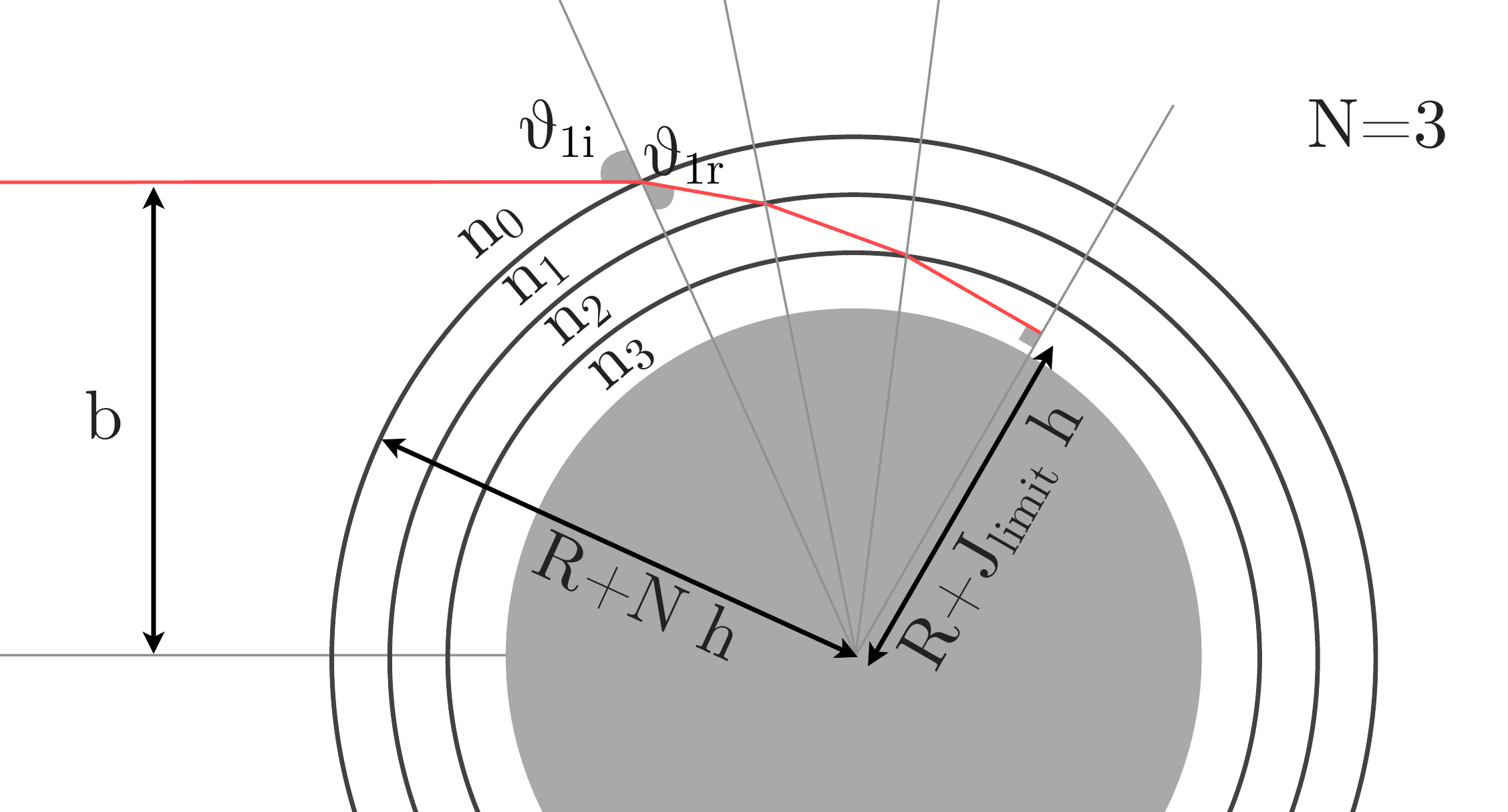}
\caption{
Schematic of a $N=3$ shell atmosphere where (exagerrated) refraction is
calculated by considering the interactions at each shell boundary.
}
\label{fig:triplelayer}
\end{center}
\end{figure*}

Consider a luminous source located at a large distance from the Earth
such that it can be approximated as a point source (we leave consideration of
diffuse sources of emission to future work). Light from the source
arrives at the Earth as an approximately plane-parallel wave with a wavelength
$\lambda$ and can be represented by a sum of parallel rays, each with an impact
parameter $b$ (see Figure~\ref{fig:triplelayer}).
The Earth's atmosphere is considered to be
described by a one-dimensional temperature-pressure profile and is unchanging
in time. The Earth's atmosphere is then split up into a series of $N$ shells,
within which the temperature, pressure and refractivity is assumed to be
constant.

The total geopotential altitude of the atmosphere is defined as $Z$, such that each
shell has a thickness of $h=Z/N$ and a refraction index of $n_j$. Shell indices are
assigned in ascending order such that the deepest layer have the highest index. The
outer shell, shell $j=0$, effectively extends out to infinity and has a refraction
index of $n_0=1$.

When the ray of light reaches the boundary between from shell $j-1$ to $j$, the
change in atmospheric density leads to a change in the light's speed and thus
refraction occurs as described by Snell's law

\begin{align}
n_{j-1} \sin \theta_{i,j} &= n_j \sin \theta_{r,j},
\end{align}

where the indices $i$ and $r$ refer to ``incidence'' and ``refraction''. The
deflection angle of the ray is therefore

\begin{align}
\alpha_j = \theta_{i,j} - \theta_{r,j}.
\label{eqn:deflection}
\end{align}

At the boundary of $j=1$ shell, the angle of incidence can be written in terms
of the impact parameter, $b$, by the trigonometric relation

\begin{align}
\sin \theta_{i,1} &= \frac{b}{1} \frac{1}{R + N h}
\end{align}

and thus via Snell's law we also have

\begin{align}
\sin \theta_{r,1} &= \frac{b}{n_1} \frac{1}{R + N h}.
\end{align}

The angle of incidence at the boundary of the $j=2$ shell can be deduced from
this result, by applying the sine rule inside the triangle subtended from
the $j=1$ boundary intersection to the $j=2$ boundary intersection to the
Earth's centre:

\begin{align}
\sin \theta_{i,2} &= \frac{R+ N h}{R + (N-1) h} \sin \theta_{r,1},\nonumber\\
\qquad&= \frac{b}{n_1} \frac{1}{R + (N-1) j}
\end{align}

And again using Snell's law this gives

\begin{align}
\sin \theta_{r,2} &= \frac{b}{n_2} \frac{1}{R + (N-1) j}.
\end{align}

Continuing this process, it is easy to show that

\begin{align}
\sin \theta_{i,j} &= \frac{b}{n_{j-1}} \frac{1}{R + (N-j+1) h}.
\label{eqn:incidence}
\end{align}

and

\begin{align}
\sin \theta_{r,j} &= \frac{b}{n_j} \frac{1}{R + (N-j+1) h}.
\label{eqn:refraction}
\end{align}

Using Equations~(\ref{eqn:incidence}) \& (\ref{eqn:refraction}) with
Equation~(\ref{eqn:deflection}) allows one to calculate the deflection
angle at each shell boundary. In practice, this is done by using the
sine addition rule

\begin{align}
\sin \alpha_j &= \sin(\theta_{i,j} - \theta_{r,j}),\nonumber\\
\qquad &= \sin \theta_{i,j} \cos\theta_{r,j} - \sin\theta_{i,j}\cos\theta_{i,j},\nonumber\\
\qquad &= \sin \theta_{i,j} \sqrt{1-\sin^2\theta_{r,j}} - \sin\theta_{i,j}\sqrt{1-\sin^2\theta_{i,j}}
\end{align}

such that

\begin{align}
\alpha_j =& \sin^{-1}\Bigg[ \nonumber\\
&\frac{b}{n_{j-1}} \frac{1}{R + (N-j+1) h} \sqrt{1 - \Big(\frac{b}{n_j} \frac{1}{R + (N-j+1) h}\Big)^2 }- \nonumber\\
\qquad& \frac{b}{n_j} \frac{1}{R + (N-j+1) h} \sqrt{1 - \Big(\frac{b}{n_{j-1}} \frac{1}{R + (N-j+1) h}\Big)^2}  \Bigg].
\end{align}

\subsection{Critical refraction limits}
\label{sub:critrefraction}

If the impact parameter is low enough, the ray will eventually strike the
planetary surface and thus end its journey. Consider the critical case of this
such that the ray just grazes the planetary surface tangentially, initiated
from an impact parameter $b_{\mathrm{crit}}$. If this happens, then it follows
that the ray must also reach the deepest atmospheric shell. In that shell, one
can draw a right-angled triangle from the planet's center to the two
intersection points and write that

\begin{align}
\sin \theta_{r,N} &= \frac{R}{R+h}
\end{align}

Now substituting in Equation~(\ref{eqn:refraction}), we have

\begin{align}
\frac{b_{\mathrm{crit}}}{n_N} \frac{1}{R + h} &= \frac{R}{R+h},
\end{align}

which may be solved to give

\begin{align}
b_{\mathrm{crit}} \equiv R n_N.
\end{align}

If the impact parameter is in the range $b_{\mathrm{crit}} < b < R + N h$, then
the ray will penetrate the Earth's outer atmospheric shell and continue down
to some depth before making its way back out of the atmosphere again. Let us
write that the deepest shell is given by the index $J_{\mathrm{limit}}$.
In order to calculate the total deflection angle, $\Delta$, of a ray, we must
calculate $J_{\mathrm{limit}}$ since $\Delta = \sum_{j=1}^{J_{\mathrm{limit}}}
\alpha_j$.

This may be calculated by considering that the $\cos\theta_{i,j}$ term in the
$\Delta$ calculation must be real. The term becomes imaginary if
$\sin\theta_{i,j}>1$. Using Equation~(\ref{eqn:incidence}), one sees that this
corresponds to

\begin{align}
\frac{1}{n_{j-1}} \frac{b}{R+(N-j+1) h} > 1.
\end{align}

Therefore, one may find $J_{\mathrm{limit}}$ by sequentially calculating the
above inequality from $j=1$ up until the condition holds true. At this point,
the previous shell is assigned as $J_{\mathrm{limit}}$. The total deflection
angle, from the lowest altitude to space (space-to-space is simply twice
as much) is then computed as

\begin{align}
\Delta = 2\sum_{j=1}^{J_{\mathrm{limit}}} \alpha_j.
\label{eqn:totaldeflection}
\end{align}

\subsection{Airmass}
\label{sub:airmass}

Aside from computing the deflection angle and deepest shell layer of an
incoming ray, one can also compute the airmass traversed, $X$. The path
length can be found by using the sine rule inside the triangles formed
between the planet's center and the shell intersection points:

\begin{align}
d_j &= \Big(\frac{ \sin(\theta_{i,j+1}-\theta_{r,j})}{\sin\theta_{i,j+1}}\Big) \big(R + (N-j+1)h\big).
\label{eqn:path}
\end{align}

Substituting Equation~(\ref{eqn:incidence}) \& (\ref{eqn:refraction}) into
Equation~(\ref{eqn:path}), and after simplification, yields

\begin{align}
d_j =& \sqrt{ (R + (N-j+1)h)^2 - (b/n_j)^2 } \nonumber\\
\qquad& - \sqrt{ (R + (N-j)h)^2 - (b/n_j)^2 }.
\label{eqn:pathlength}
\end{align}

The airmass passed through by a ray is proportional to the path length
multiplied by the density. Using the ideal gas law, the density is
proportional to pressure over temperature ($P/T$). Accordingly, airmass
is given by

\begin{align}
X &= \frac{\sum_{j=1}^{J_{\mathrm{limit}}} d_j P_j/T_j}{X_0},
\end{align}

where $X_0$ is a constant of proportionality defined such that
$X=1$ for a ray which travels from sea level to space along the
zenith. This equates to

\begin{align}
X_0 \equiv \sum_{j=1}^N h P_j/T_j.
\end{align}

\subsection{Model atmosphere}
\label{sub:modelatmosphere}

As stated earlier, this work assumes that within a given shell, the pressure,
temperature and refractivity are constant. In other words, a one-dimensional
static atmosphere. The purpose of this work is to simply demonstrate the
concept of the terrascope. If worthwhile, future work could be undertaken to
use more sophisticated atmospheric models accounting for weather, turbulence,
and regional differences. For now, the goal is merely
to compute the approximate feasibility and properties of the terrascope, by
making reasonable but ultimately simplifying assumptions.

\begin{figure}
\begin{center}
\includegraphics[width=8.4cm,angle=0,clip=true]{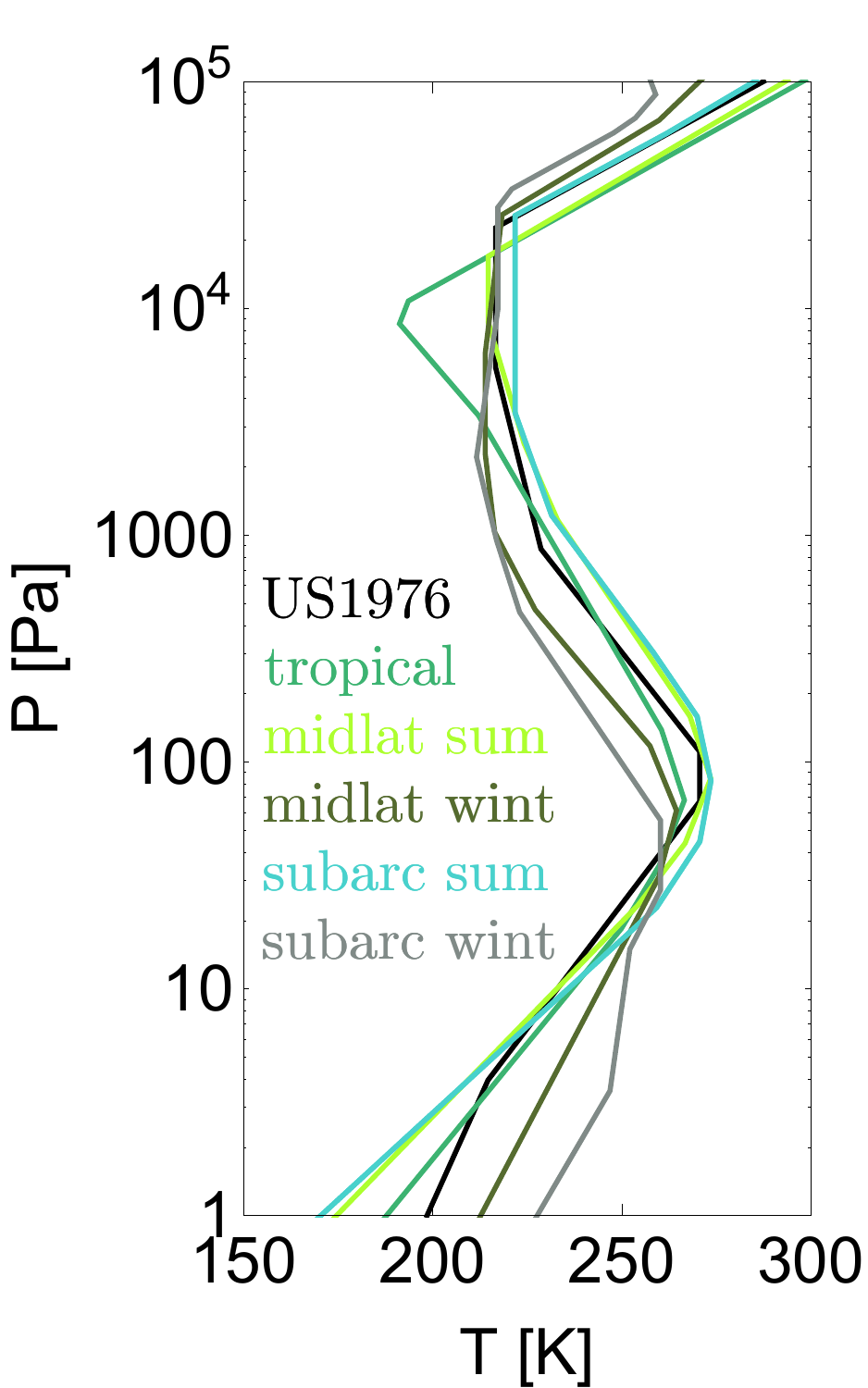}
\caption{
Temperature-pressure profiles for six standard atmospheres used in
this work.
}
\label{fig:TP}
\end{center}
\end{figure}

To accomplish this, the US Standard Atmosphere 1976 \citep{US:1976} was adopted
as a fiducial temperature-pressure (TP) profile. This atmosphere can be
considered to be an average over the global climate but can be a poor
representation for particular local climates. To investigate the impact of
differing conditions, five other standard TP profiles were utilized - in
particular the same atmospheres as used by \lowtran\ \citep{lowtran:1988}.
These are the ``tropical'', ``mid-latitude summer'',
``mid-latitude winter'', ``sub-arctic summer'' and ``sub-arctic winter''
models (these are shown in Figure~\ref{fig:TP}).

The models define a continuous temperature-pressure profile from 0 to
85\,km geopotential altitude ($z$) but with functional changes occurring
at six key boundaries distributed in altitude. Within each of the
layers (defined by sharp boundaries), the lapse rate, $\mathbb{L}$, is
varied and the pressure computed assuming an ideal gas and vertical
pressure variation of $\mathrm{d}P/\mathrm{d}z = -\rho g$ (where
$\rho$ is the density of air and $g$ is the acceleration due to gravity).

Temperature, as a function of geopotential altitude, within the 
$k^{\mathrm{th}}$ layer is defined by

\begin{equation}
T_k[a] = T_{k-1} + \mathbb{L}_k (z-z_k),
\end{equation}

where $T_0$ is the temperature at sea level. The pressure is then given by

\begin{equation}
P_k[z] =
\begin{cases}
P_{k-1} \exp\Big( -\frac{-g_0 M (z-z_k)}{R T_{k-1}} \Big)  & \text{if } \mathbb{L}_k=0,\\
P_{k-1} \Big(\frac{T_{k-1}}{T_k[z]}\Big)^{ \frac{g_0 M}{R \mathbb{L}_k} }  & \text{otherwise},\\
\end{cases}
\end{equation}

where $P_0$ is the pressure at sea level and $g_0 M/R = 34.1645$\,K/km.

\subsection{Calculating refractivity}
\label{sub:refractivity}

The refractivity of air, $\eta$, equals the refraction index, $n$, minus unity.
Given a shell's pressure and temperature defined by the US Standard Atmosphere
1976, the refractivity may be computed using a semi-empirical formula. In this
work, the expression of \citet{birch:1994} is adopted since the expression
is found to be in better agreement with recent measurements than the
\citet{edlen:1966} formula - largely due to the increase in ambient carbon
dioxide levels \citep{birch:1994}. The refractivity of dry air is thus given by

\begin{align}
\eta = 10^{-8} P &\Big( C_1 + \frac{C_2}{C_3-\sigma^2} + \frac{C_4}{C_5-\sigma^2} \Big) \nonumber\\
\qquad& \Big( \frac{ 1 + 10^{-10}P (C_6-C_7 T')}{ C_8 (1+C_9 T') } \Big),
\end{align}

where $T'$ is the temperature in Celsius, $\sigma$ is the reciprocal
of the wavelength of light in a vacuum in units of nanometres, and
$C_1$ to $C_9$ are constants.

The calculations described throughout the rest of the paper were also repeated
using moist air refractivity, instead of dry air, using the appropriate correction
\citep{birch:1994}. However, this was found to produce very minor changes to the
results and thus the dry air formula will be used in what follows.

It is instructive to consider the approximate relationship as well. Refractivity is
proportional to the gas density. For an isothermal atmosphere, one expects
$\rho \propto e^{-z/H}$, where $H$ is the scale height and $z$ is the altitude.
Accordingly, one expects $\eta \propto e^{-z/H}$ too.

\section{Ray Tracing Simulations}
\label{sec:raytracing}

\subsection{Generating a training set}
\label{sub:trainingset}

Two physical principles are critical in consideration of the terrascope,
refraction and extinction. The issue of atmospheric extinction will be tackled
later in Section~\ref{sub:onaxisextinction}, and thus we first deal with
refraction using the expressions found earlier in Section~\ref{sec:refraction}
to ray trace through the Earth's atmosphere. In what follows, rays are only
traced from space to the point of closest approach to the Earth's surface.
Since the assumed model atmosphere is static and one-dimensional, the egress
path will be identical to the ingress path and one may exploit this symmetry
to save computational effort.

Before a ray can be traced, it is first necessary to choose how many shells
($N$) will be used for ray tracing experiment. In general, the greater the
number of shells, the more accurate the integration, but at greater computational
expense. Further, it is necessary to choose up to what geopotential altitude
the atmosphere terminates, $Z$ (technically the atmosphere extends to infinity
but this of course is not computationally reasonable, nor are the standard
atmospheres well-defined above \SI{86}{\kilo\metre}).

In preliminary ray tracing experiments, it was noted the deflection angles
follow a generally smooth trend with respect to impact parameter until the
impact parameter approaches $R$ + 86\,km. At the 86\,km boundary, the
refractivity is discontinuous, jumping off a monotonically decreasing
smooth function down zero. This appears to introduce peculiar behavior for
extreme impact parameters and thus $Z=$\SI{80}{\kilo\metre} was adopted in an
effort to avoid this regime. A high but computationally efficient resolution
was chosen such that each shell has a thickness of $h=$\SI{10}{\centi\metre},
corresponding to 800,000 steps.

Let us choose a particular wavelength, $\lambda$, of light to work with. Using
this wavelength, one still needs to choose an impact parameter, $b$, before
ray tracing can be executed. To generate a suite of examples, let us define a
grid of impact parameters varying from $R$ to $R+Z$ uniformly spaced in
geopotential altitude. Since the numerical resolution of the atmosphere is
itself 800,000, the resolution used for $b$ cannot be higher than this and
a reasonable choice is to adopt an order-of-magnitude lower to ensure the
highest possible scan yet retain reliable results. Accordingly, a
resolution in $b$ of 80,000 was adopted (i.e. \SI{1}{\metre} step sizes).

These experiments essentially generate a training set from which one can learn
the relationship between $b$ and deflection angle, $\Delta$ (as well as
other properties). However, the set is conditioned upon a specific choice
of $\lambda$. To build a complete training set, it is necessary to also vary
$\lambda$. This is done by creating a grid from
$\lambda=$\SI{0.2}{\micro\metre} to \SI{30}{\micro\metre} with 219 examples
spaced log-uniformly. This wavelength range corresponds to the range of support
for the \lowtran\ extinction model that will be used later (see
Section~\ref{sub:onaxisextinction}).

In each run, the total deflection angle, $\Delta$, is recorded, as well as
the minimum geopotential height achieved by the ray\footnote{
In this way, $b$ and $D$ are closely related; $b$ is the minimum separation
between the Earth's center and the ray in the absence of refraction,
whereas $D$ is the same but with refraction turned on.
} (``depth''), $D$, and
the airmass traversed through, $X$. It was found that impact parameters close
to the \SI{80}{\kilo\metre} boundary exhibited slight trend differences than
the bulk, suggestive of a numerical error. To alleviate this, samples with
$b>$\SI{77}{\kilo\metre} were excluded, as well as samples for which the ray
strikes the Earth, leaving a total of 10,606,382 ray tracing experiments that
were saved.

\subsection{Interpolation scheme}
\label{sub:interpolation}

In order to generalize the numerical results to arbitrary values of $b$
and $\lambda$, one can apply interpolation to the training data.

\subsubsection{Critical impact parameter, $b_{\mathrm{crit}}$}

For each ray tracing experiment, only $X$, $D$ and $\Delta$ are saved
and so these are the three terms that require interpolation. However,
a useful product of these is $b_{\mathrm{crit}}$, the impact parameter
at which $D=R$. Since the simulations iterate through in 1\,metre steps
in $b$, one may simply cycle through the list until $D<R$ and save the
previous example as $b_{\mathrm{crit}}$, which will have a maximum
associated error of \SI{1}{\metre}.

\begin{figure}
\begin{center}
\includegraphics[width=1.05\columnwidth,angle=0,clip=true]{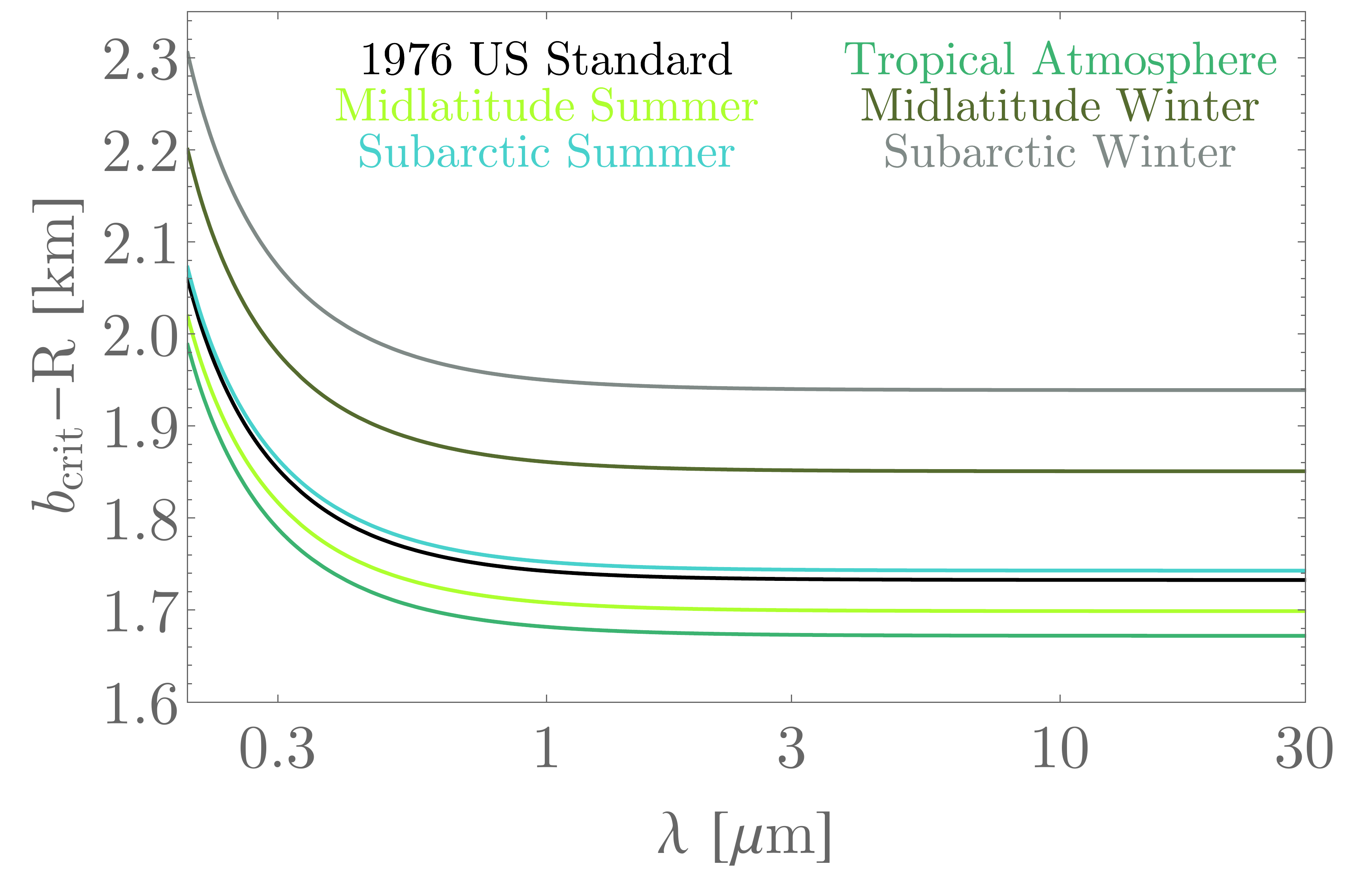}
\caption{
The critical impact parameter as a function of wavelength. Impact
parameters below this will refract so much they strike the Earth.
The different lines shows the effect of varying the climate model.
}
\label{fig:bcrit}
\end{center}
\end{figure}

In total, there are 220 training examples of $b_{\mathrm{crit}}$ versus
$\lambda$. When cast as $b_{\mathrm{crit}}$ against $\lambda^{-2}$,
the relationship appears quasi-linear and thus it is using this
parameterization that the interpolation is performed. Since the number
of samples is relatively small, it is feasible to perform Gaussian
process (GP) regression \citep{gp1,gp2}, in this case using a squared exponential kernel.
Using exhaustive leave 1-out, this process is repeated to evaluate the
error in the final estimates. The final interpolative function, shown in
Figure~\ref{fig:bcrit}, has a maximal error of one metre, which is the
numerical error of the training set to begin with. It therefore represents
an excellent predictor and is adopted in what follows.

\subsubsection{Airmass, $X$, and depth, $D$}

\begin{figure}
\begin{center}
\includegraphics[width=1.05\columnwidth,angle=0,clip=true]{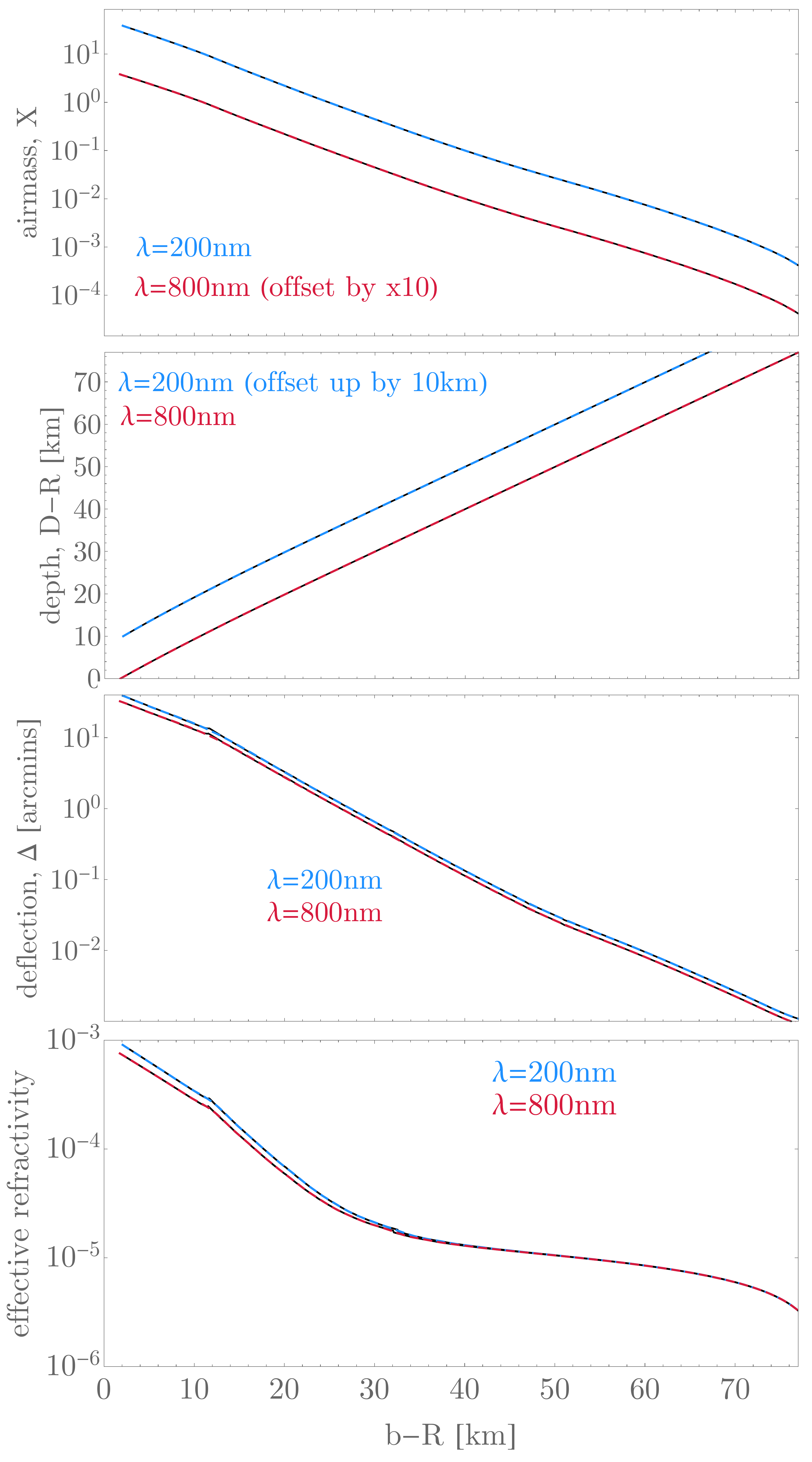}
\caption{
Simulation results are shown in dark gray, interpolative functions
in dashed colors.
Upper: Airmass traversed for a ray traveling through the Earth's
atmosphere from space to its closest approach to the Earth as
function of impact parameter.
Middle-upper: Depth of the ray at closest approach.
Middle-lower: Deflection angle of the ray by the time it reaches closest
approach (rays interior to the critical impact parameter are omitted).
Lower: ``Effective'' refractivity of the atmosphere, calculated
as described in the main text.
}
\label{fig:interp}
\end{center}
\end{figure}

For airmass and depth, a Gaussian process regression is impractical due to the
much larger and two-dimensional training set of over 10 million samples.
Instead, this large sample is suitable for a dense interpolative net.
As with $b_{\mathrm{crit}}$, it was found that $\lambda^{-2}$ provides a
more linear basis for training for both $X$ and $D$. The 2D bilinear
interpolation therefore maps $\{\lambda^{-2},b\} \to X$ and $D$.
Examples of the interpolative function are shown in Figure~\ref{fig:interp}.

To provide further intuition and context, we also show the ``effective''
refractivity of the Earth's atmosphere in Figure~\ref{fig:interp}. This is
computed by taking the computed deflection angles and solving for the
equivalent refractivity needed for that deflection using a single-layer
atmosphere.

Using leave 1-out, one may evaluate the error of these interpolations. This
is done by leaving a random example out, re-training the interpolation,
and then evaluating the residual between the omitted sample and the
interpolated prediction of said point. Since the training is relatively
expensive computationally, the aforementioned Monte Carlo experiment is
limited to 1000 realizations.

This process revealed the bilinear scheme is able to perform excellent
interpolation across the grid. The relative airmass residuals show a
non-Gaussian distribution with a standard deviation of 0.68\%, a mean
absolute error of 0.13\% and a 99.9\% of samples exhibiting an absolute
error less than 0.11\%. For depth, the standard deviation of the
absolute residuals is 0.030\,m, the mean absolute error is 0.012\,m,
and 99.9\% of samples have an absolute error less than 0.20\,m.

\subsubsection{Deflection angle, $\Delta$}

For deflection angle, it was found that simple bilinear interpolation did not
provide particularly stable results, too closely tracing out the small
numerical errors found in the simulations rather than smoothing over them.
Another problem with this scheme was that as the simulations approach $b
\to b_{\mathrm{crit}}$, there is no training data for deflection
angle since it cannot be defined below this point. This led to unstable
extrapolations in the final shell.

Instead of bilinear interpolation, a Gaussian process with a rational quadratic
kernel is trained on each wavelength slice independently. The training data
is also thinned by a factor of 100 to expedite the training (leaving
approximately 750 samples per slice). Rather than use $\Delta$ as the target
function, we use $\log \Delta$ which behaves quasi-linearly with respect to
$(b-R)$, the independent variable for the training. The GP is used because it
smooths over the numerical noise introduced by our machine-precision
calculations.

Calling the GP predictor is computationally slow, and so a library of
predictions is generated for later use. This library samples the original
simulations at a thinning rate of ten along the $b$-axis, giving 7506 samples.
The library is then interpolated using splines in cases where needs to evaluate
the deflection angle at intermediate choices of $\Delta$. Comparing to the
original samples, the agreement of the final interpolations is better than
1.9\% for all samples, with a standard deviation of 0.83\%. The interpolated
function is shown in Figure~\ref{fig:interp}.

\subsection{Computing on-axis amplification}
\label{sub:onaxis}

Amplification is defined here as the intensity received by a detector using the
terrascope relative to the intensity the same detector would receive in the
absence of the Earth. The latter of these two terms is simply the incident
flux multiplied by the collecting area of the detector, $\pi (W/2)^2$, where
$W$ is the diameter/aperture of the detector. In the same way, the intensity
received by the terrascope can be computed by simply considering the effective
light-collecting area. In what follows, it is assumed that the source, the lens
and the detector are all perfectly aligned, which is referred to as ``on-axis''.

\begin{figure*}
\begin{center}
\includegraphics[width=17.0cm,angle=0,clip=true]{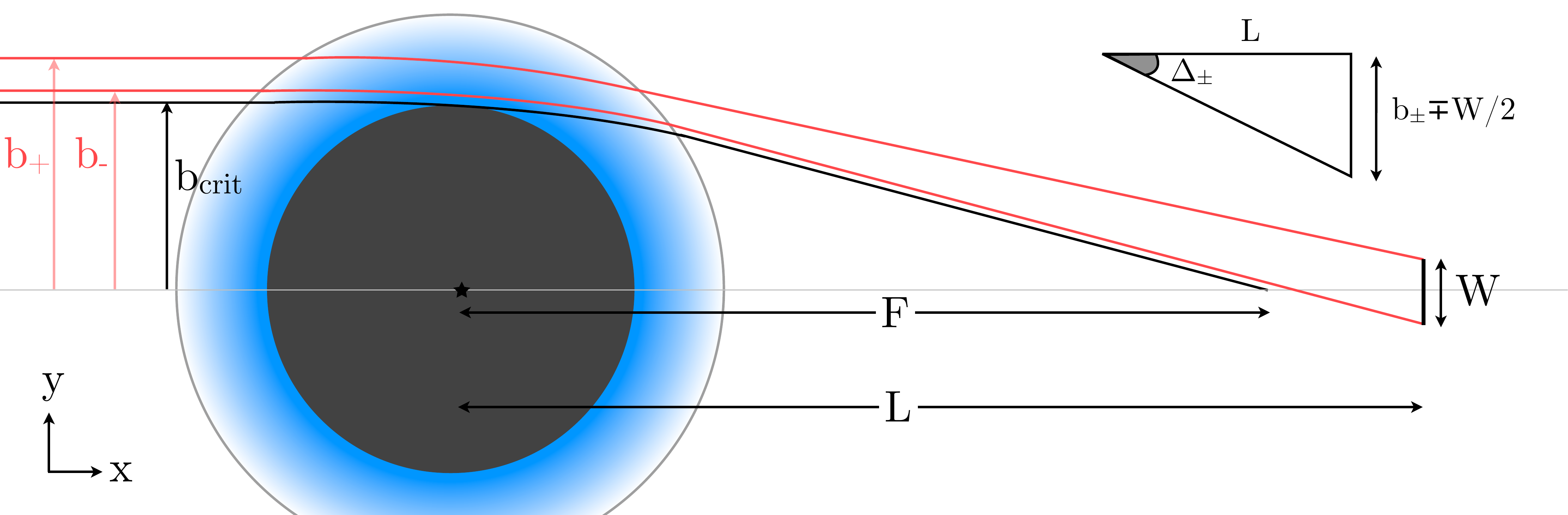}
\caption{
Illustration of a detector of diameter $W$ utilizing the terrascope. Two rays
of different impact parameters, but the same wavelength, lens through the
atmosphere and strike the detector. The ring formed by those two rays
enables a calculation of the amplification. In this setup, the detector
is precisely on-axis.
}
\label{fig:onaxis}
\end{center}
\end{figure*}

The setup is illustrated in Figure~\ref{fig:onaxis}. A ray of wavelength
$\lambda$ and impact parameter of $b_{-}$ is refracted by an angle
$\Delta_{-}$ such that it strikes lower tip of the detector located at a
distance of $L$. Another ray with the same wavelength but a higher impact
parameter, $b_{+}$, is refracted by an angle $\Delta_{+}<\Delta_{-}$ and
eventually strikes the upper tip of the detector. It follows that all rays of
wavelength $\lambda$ and impact parameter $b_{-} \leq b \leq b{+}$ will strike
the detector. In the on-axis case considered here, along with the assumption
of a 1D atmosphere, the problem is symmetric about the $x$-axis and thus
the lensing region is circular ring of area $\pi (b_{+}^2 - b_{-}^2)$,
meaning that the amplification $\mathcal{A}$, is given by

\begin{align}
\mathcal{A} &= \epsilon \frac{b_{+}^2 - b_{-}^2}{(W/2)^2},
\label{eqn:ampdefinition}
\end{align}

where $\epsilon$ is a loss parameter describing the degree of extinction.
Rather than forming a single focus point, light focusses along a line
much like the case of gravitational lensing. The maximum distance of the
focal line is infinity, but the inner distance is well-defined and
it is labelled as $F$ in what follows. This distance corresponds to
a ray striking the Earth at the critical impact parameter,
$b_{\mathrm{crit}}$ (for a given wavelength). The focal distance is
given by simple trigonometry

\begin{align}
F &= b_{\mathrm{crit}} \cot \Delta_{\mathrm{crit}}
\end{align}

where

\begin{align}
\Delta_{\mathrm{crit}} &\equiv \lim_{b \to b_{\mathrm{crit}}} \Delta(b).
\end{align}

\begin{figure}
\begin{center}
\includegraphics[width=1.05\columnwidth,angle=0,clip=true]{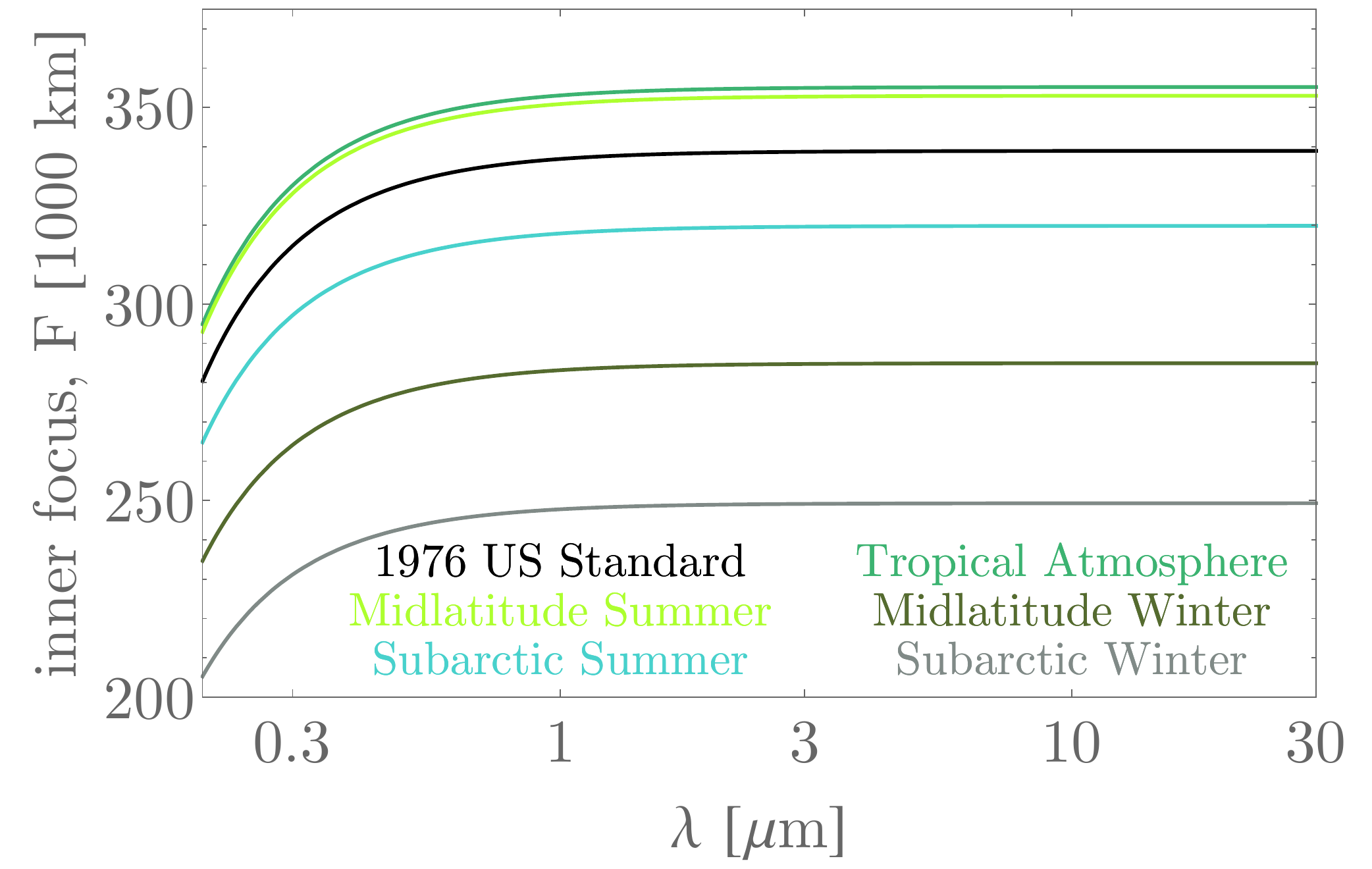}
\caption{
Location of the inner focal point of the terrascope as a function of
wavelength. Rays cannot focus interior to this point because they would strike
the Earth's surface. Results shown for six different model temperature-pressure
profiles.
}
\label{fig:focus}
\end{center}
\end{figure}

In the wavelength range of \SI{0.2}{\micro\metre} to \SI{30}{\micro\metre}, the
inner focus point varies from ${\simeq}$\SI{200000}{\kilo\metre} to
${\simeq}$\SI{350000}{\kilo\metre}, depending on the wavelength and climate
model (see Figure~\ref{fig:focus}). This indicates that it would be possible to
focus light at the lunar distance itself since the focal line extends to
infinity past this inner point. Accordingly, observatories at or beyond the
lunar distance could be feasible locations for the terrascope detector.

The impact parameters $b_{+}$ and $b_{-}$ dictating the amplification can be
derived by geometrical arguments. Consider a ray of impact parameter $b_{+}$
which deflects by angle $\Delta[b_{+}]$ such that it strikes the upper tip of
the detector located at a distance $L$. Since the offset from the $x$-axis
of the upper detector tip is $W/2$ and thus one can write that

\begin{align}
b_{+} - L \tan\Delta[b_{+}] = W/2.
\label{eqn:bplus}
\end{align}

Similarly, consider a ray which passes a little deeper through the planetary
atmosphere at impact parameter $b_{-}$, such that it deflects enough to
strike the lower tip of the detector, which satisfies

\begin{align}
b_{-} - L \tan \Delta[b_{+}] = -W/2.
\label{eqn:bminus}
\end{align}

In practice, solutions for these two impact parameters are found through a
numerical Nelder-Mead optimizer, because of the subtle dependency of $\Delta$
upon $b$ (see Section~\ref{sub:interpolation}).

This optimization is repeated for various choices of $L$, $W$ and $\lambda$.
For $\lambda$, the original grid of 220 wavelengths was adopted. For $L$, it
was found that the inner focus of the shortest wavelength with the US Standard
Atmosphere 1976 model was \SI{281700}{\kilo\metre} and thus a uniform grid was
adopted from this distance out to \SI{1500000}{\kilo\metre} (the Hill sphere
of the Earth) with 101 steps. Finally, for $W$, five fiducial diameters are
adopted: $10^{-2}$, $10^{-1}$, $10^{0}$, $10^1$ and $10^2$ metres.

\subsection{Analytic estimates}
\label{sub:analytic}

Although the numerical experiments provide the most precise view, it is
instructive to consider the approximate scaling relations
expected. One can note that the $\Delta$ function is approximately log-linear
and thus can be approximated as $\Delta \simeq \Delta_0 e^{-(b-R)/H_{\Delta}}$
where $H_{\Delta}$ is an effective scale height for the lensing, equal to
\SI{6.911}{\kilo\metre} to within three-decimal places for all rays between
$\lambda=$\SI{0.2}{\micro\metre} to $\lambda=$\SI{30}{\micro\metre}. Using this
approximate formalism, Equations~(\ref{eqn:bplus}) \& (\ref{eqn:bminus}) can be
combined to give

\begin{align}
W =& \Delta b + \nonumber\\
\qquad& L \Bigg( \tan\Big(\Delta_0 e^{-(b_{-}-R)/H_{\Delta}}\Big) - \tan\Big(\Delta_0 e^{-(b_{+}-R)/H_{\Delta}}\Big) \Bigg)
\end{align}

where $\Delta b = (b_{+}-b_{-})$. Taking a small-angle approximation,
replacing $b_{-}=b_{\mathrm{mid}} - \Delta b/2$ and $b_{+}=b_{\mathrm{mid}}
+ \Delta b/2$, and then Taylor expanding for small $\Delta b$ (thin
ring approximation) gives

\begin{align}
\Delta b \Bigg( 1 + \frac{L}{H_{\Delta}} \Delta_0 e^{-\frac{b_{\mathrm{mid}}-R}{H_{\Delta}}} \Bigg) = W.
\end{align}

In order to reach the detector, one may write that $\Delta \simeq b/L$, or
simply $\Delta \sim R/L$ by noting that $R \gg (b-R)$. This allows us to write that

\begin{align}
\Delta b \sim& \frac{W}{1 + (L/H_{\Delta}) (R/L)},\nonumber\\
\qquad \sim& \frac{W}{(R/H_{\Delta})},
\label{eqn:approxthickness}
\end{align}

where the second lines has used the fact $R \gg H_{\Delta}$.
Whilst one might naively intuit that $\Delta b \sim W$, the gradient in
the refractive index means the rays needs to be closer together than
this, since even a slight angular difference is magnified over the large
distance $L$. The denominator is of order $10^3$ and thus implies that for
a one-meter diameter detector, the lensing ring is about a millimetre thick.

The above allows one to approximately estimate the amplification,
$\mathcal{A}$. If one writes that $b_{+} = b_{-} + \Delta b$, then
Equation~(\ref{eqn:ampdefinition}) becomes $\mathcal{A} = 
\epsilon b_{-}^2 ((1 + \tfrac{\Delta b}{b_{-}})^2-1)/(W/2)^2$. Since
$\Delta b \ll W$ and $b_{-} \sim R$, then provided $W \ll R$ (which practically
speaking will always be true), one may write that
$\tfrac{\Delta b}{b_{-}} \ll 1$. This permits a Taylor expansion of
$\mathcal{A}$ such that

\begin{align}
\mathcal{A} \simeq 2 \epsilon b_{-} \Delta b/(W/2)^2
\end{align}

which can be further refined by adopting $b_{-} \sim R$ and using
Equation~(\ref{eqn:approxthickness}) to write

\begin{align}
\mathcal{A} \sim& 2 \epsilon R \frac{W}{(R/H_{\Delta})} \Big(\frac{4}{W^2}\Big),\nonumber\\
\qquad \sim& 8 \epsilon H_{\Delta}/W.
\label{eqn:approxamp}
\end{align}

Since $H_{\Delta} = $\SI{6.911}{\kilo\metre}, then $\mathcal{A}/\epsilon \sim 55000/W$,
which gives a first estimate for the approximate degree of amplification
expected. The effective aperture size is given by $\sqrt{\mathcal{A}}$
and thus equals $W_{\mathrm{eff}} \sim 235 \epsilon^{1/2} \sqrt{W/(\mathrm{metres})}$\,metres.

%
%
%
%

\subsection{Computing off-axis amplification}
\label{sub:offaxis}

The alignment of the source, lens and detector is instantaneous. Whilst useful
for estimating the limiting amplification, practically speaking the source
spends infinitesimal time at this position and so the useful lensing time
is defined by the off-axis positions. Amplification still occurs off-axis
but now the rays which reach the detector must be deflected by different
angles, depending upon whether the rays travel above or below the mid-plane.

\begin{figure*}
\begin{center}
\includegraphics[width=17.0cm,angle=0,clip=true]{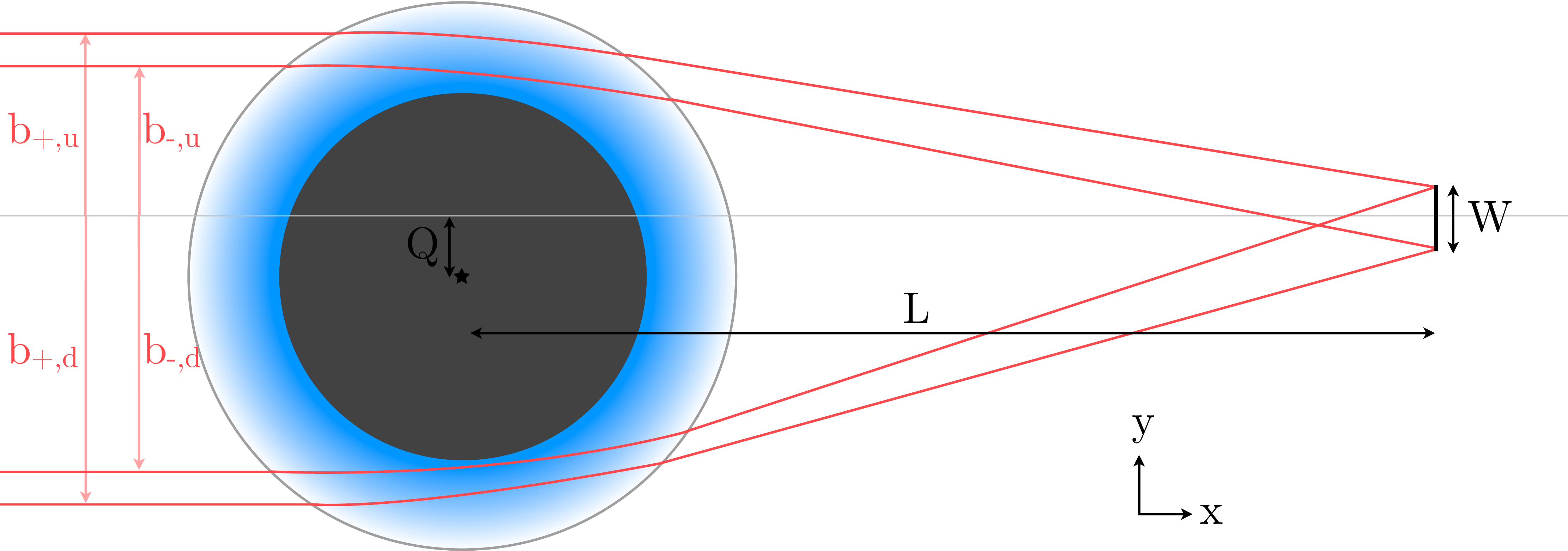}
\caption{
Same as Figure~\ref{fig:onaxis} except for off-axis lensing. Only the extrema
rays in the $z=0$ plane are shown.
}
\label{fig:offaxis}
\end{center}
\end{figure*}

Consider that the Earth is offset from the line connecting the source
and the detector's mid-point by a distance $Q$, as shown in
Figure~\ref{fig:offaxis}. Although in reality the Earth is three-dimensional
and rays can take different paths than the four lines shown in this
diagram, the four rays represent the most extreme affected paths as a
result of the translation shift. All four rays can reach the detector
provided the following four conditions hold true

\begin{align}
&L \tan \Delta[b_{-,d}] - b_{-,d} = +W/2 + Q ,\nonumber\\
&L \tan \Delta[b_{+,d}] - b_{+,d} = -W/2 + Q ,\nonumber\\
&L \tan \Delta[b_{-,u}] - b_{-,u} = +W/2 - Q ,\nonumber\\
&L \tan \Delta[b_{+,u}] - b_{+,u} = -W/2 - Q .
\end{align}

Or more generally, received rays satisfy

\begin{align}
L \tan \Delta[b_d] - b_d - Q \leq |W/2|,\nonumber\\
L \tan \Delta[b_u] - b_u + Q \leq |W/2|.
\label{eqn:updown2D}
\end{align}

Naturally, if $Q\gtrsim(R+Z)$, then no deflection is required and
rays will arrive at the detector unlensed above the detector axis.

Consider a ray which now lives out of the plane, with a $\hat{z}$-axis offset
of $\beta_{d,z}$. For a ray below the $\hat{x}$-axis, the incident ray has
a detection axis offset in the $\hat{y}$-direction of $\beta_{d,y} + x$, which
when combined with the $z$ term gives a Euclidean offset from the
detector axis of $\beta_d=\sqrt{(\beta_{d,y} + Q)^2 + \beta_{d,z}^2}$ but an
impact factor from the planet of $b_d=\sqrt{\beta_{d,y}^2 + \beta_{d,z}^2}$.
One may write that $\beta_{d,y} = b_d \cos \phi$ and
$\beta_{d,z} = b_d \sin \phi$, where $\phi$ is the azimuthal angle about the
$\hat{x}$-axis of the incident ray. Now the total offset from the detector axis
is given by $\beta_d=\sqrt{b_d^2+Q^2+2 b_d Q \cos\phi}$. Similarly, if the
incident ray were above the detector axis, then
$\beta_u=\sqrt{b_u^2+Q^2-2 b_u Q \cos\phi}$. For a circular detector of
radius $W/2$, rays will strike the detector if

\begin{align}
L \tan \Delta[b_d] - \sqrt{b_d^2+Q^2+2 b_d Q \cos\phi} \leq |W/2|,\nonumber\\
L \tan \Delta[b_u] - \sqrt{b_u^2+Q^2-2 b_u Q \cos\phi} \leq |W/2|.
\end{align}

If $Q=0$ in the above (i.e. on-axis), then these expressions are identical to
the previous equations in Section~\ref{sub:onaxis}. Also, setting $\phi=0$
recovers Equation~(\ref{eqn:updown2D}). Accordingly, one twists $\phi$ in
the range $-\pi/2<\phi<\pi/2$ for both expressions and expects them to meet
at the extrema (which is indeed true). Moreover, one can generalize the
above pair into a single expression where $-\pi/2<\phi<3\pi/2$:

\begin{align}
L \tan \Delta[b] - \sqrt{b^2+Q^2+2 b x \cos\phi} \leq |W/2|,\nonumber\\
\end{align}

\begin{figure*}
\begin{center}
\includegraphics[width=17.0cm,angle=0,clip=true]{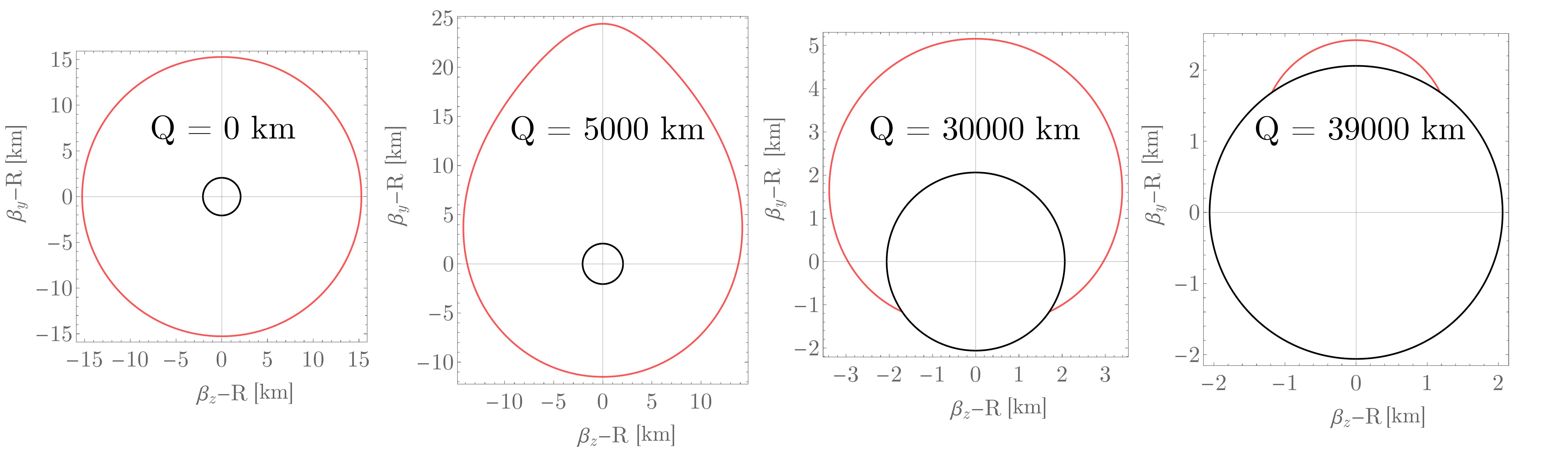}
\caption{
Numerically computed shapes of the lensing strata for three different offsets
(red lines).These are the altitudes of the rays above the Earth in order for
them to come to a focus point at distance $L$. Black lines show the critical
impact parameter inside which rays strike the planet. They represent
a kind of refractive surface, below which rays will eventually the intercept
the physical surface - which is located at the origin. Calculations use the
US Standard Atmosphere 1976, $\lambda=$\SI{0.2}{\micro\metre} and
$L = R_{\mathrm{Hill}}$. Shapes are exaggerated by virtue of the subtraction
of $R$ off both axes.
}
\label{fig:shapes}
\end{center}
\end{figure*}

Unlike the on-axis case, the ring of lensed light no longer forms a
circle and more closely resembles an egg-shape, which is illustrated in
Figure~\ref{fig:shapes}. This occurs because each lensed ray now
requires a different deflection angle to reach the detector, as a result
of the offset between the source, lens and detector. This, in turn,
means that the lensing depth - which strongly controls the deflection
angle (see Figure~\ref{fig:interp}) - is different for each lensed ray.
For $Q \ll R$, the shape is essentially circular,
for $Q \sim R$ the shape is highly oval, and for $Q \gg R$, up to a
maximum critical point, the shape disappears behind the planet.

\section{Calculation Results}
\label{sec:results}

\subsection{Aperture scaling}
\label{sub:onaxisaperture}

For a one-metre diameter telescope, typical non-extincted amplifications are
found in the range of 50,000 to 80,000 - using the numerical methods described
in Section~\ref{sec:raytracing}. For a one-metre aperture, the lensing
ring is just over a millimetre in thickness. For other aperture sizes
the thickness is found to scale with the inverse of the aperture diameter
(see Figure~\ref{fig:ampplot}) i.e. $\Delta b = (b_{+} - b_{-}) \propto W$.
These numerical results agree with the approximate analytic estimates
deduced earlier in Section~\ref{sub:analytic}.

Changing the telescope aperture has a dramatic impact on the amplification.
A clear pattern is that the amplification scales as $1/W$. For example, the
amplification of a 10-metre detector is 10 times less i.e. 7,000 to 8,000.
This result was also found in our earlier approximate analytic estimates
in Section~\ref{sub:analytic}. This scaling result indicates that one may
simply consider the results for a fixed fiducial detector and scale
appropriately. In what follows, $W=$\SI{1}{\metre} is adopted
and thus the amplification of such an aperture is denoted as $\mathcal{A}_0$.

Taking the amplification and the aperture size used, one can estimate what
the effective aperture of the telescope would have to be to match the
terrascope. Using the scaling law just described, this allows the effective
aperture to be compactly expressed as

\begin{align}
\Big(\frac{W_{\mathrm{eff}}}{\mathrm{metres}}\Big) &= \sqrt{ \mathcal{A}_0 \epsilon \Big(\frac{W}{\mathrm{metres}}\Big) }
\end{align}

This reveals that the effective aperture of the terrascope equals the actual
aperture when $W = \mathcal{A}_0$\,metres i.e. ${\sim}$\SI{80}{\kilo\metre},
setting an upper limit for the useful size of a terrascope observatory.

\subsection{Distance dependency}
\label{sub:onaxisdistance}

\begin{figure}
\begin{center}
\includegraphics[width=1.05\columnwidth,angle=0,clip=true]{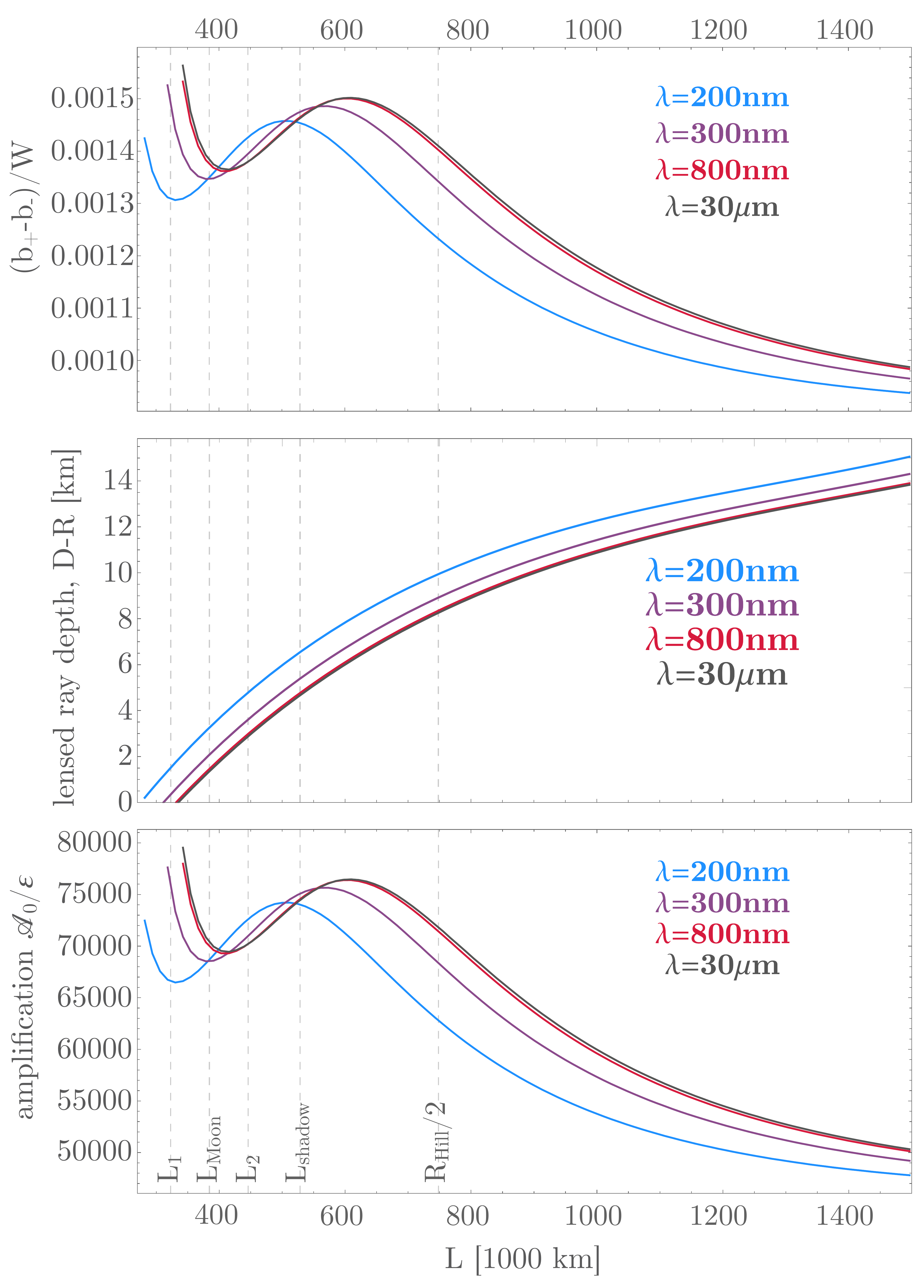}
\caption{
Amplification (lower panel) of 1\,metre detector using the terrascope
as a function of separation from the Earth, $L$, for four different
wavelengths of light. The upper panel shows the corresponding depth
of the ray, and the middle panel shows the lensing ring width.
}
\label{fig:ampplot}
\end{center}
\end{figure}

Figure~\ref{fig:ampplot} shows the amplification as a function of $L$,
illustrating how there is an overall drop-off in amplification away from the
inner focus. Although the overall maxima occurs at the inner focus, a curious
second maxima occurs at around $L=$\SI{500,000}{\kilo\metre} but appears highly
chromatic. These maxima all corresponds to rays with a depth of
${\simeq}H_{\Delta}$ revealing their commonality.

\begin{figure}
\begin{center}
\includegraphics[width=1.05\columnwidth,angle=0,clip=true]{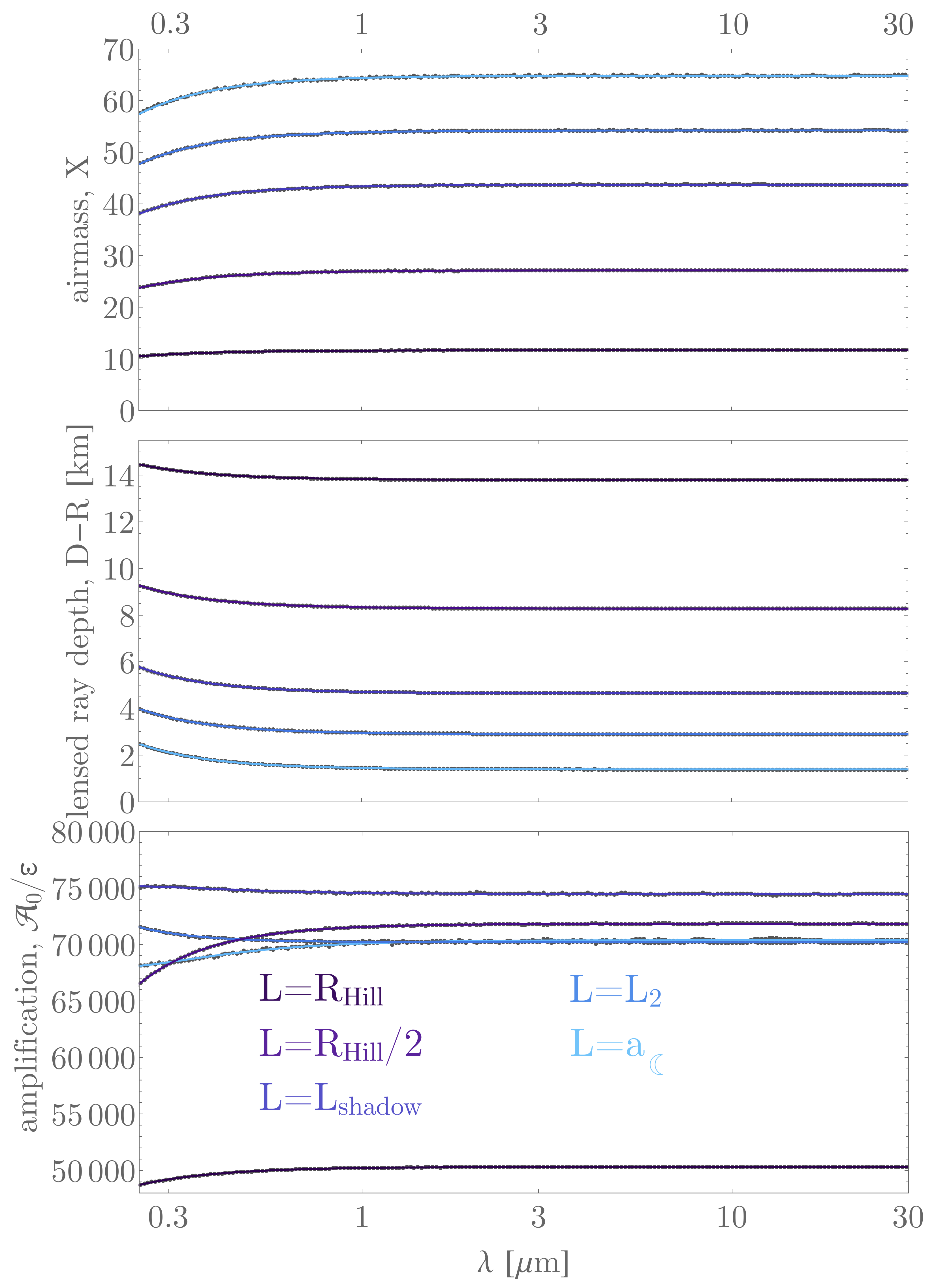}
\caption{
The airmass traversed, telluric depth and amplification as a
function of wavelength for a 1\,metre telescope at five possible
locations.
}
\label{fig:lunar}
\end{center}
\end{figure}

Consider fixing $L$ to several plausible options as depicted in
Figure~\ref{fig:lunar}, which shows the wavelength dependent amplification
for a one-metre aperture. The deepest telluric depth of the rays
received by the detector is shown in the second panel of that figure,
illustrating how redder light needs to travel deeper to reach the
observatory. The airmass traversed is shown in the top panel.

In all cases, the rays travel through a substantial amount of airmass
and thus one might question whether atmospheric extinction would
overwhelm any gains made by the terrascope setup. Two forms of
extinction are considered here - clear-sky scattering and interception
with clouds. These are dealt with separately in what follows.

\subsection{Clear-sky extinction}
\label{sub:onaxisextinction}

To estimate extinction, the \lowtran\ transmittance and radiance package is
used \citep{lowtran:1988}. Practically speaking, the code used is a \python\
wrapper implementation of \lowtran\ (available at \scivision), where the
{\tt TransmittanceGround2Space.py} script is run setting the zenith angle to
$90^{\circ}$. \lowtran\ computes transmittance from the UV/optical out
to \SI{30}{\micro\metre} and thus this defines the wavelength range considered
inn what follows.

The code is run for 41 choices of observer height, from \SI{0.01}{\kilo\metre}
to \SI{100}{\kilo\metre} in log-uniform steps. The \SI{100}{\kilo\metre} run is
so close to 100\% transmittance it is defined as such in what follows to
provide a crisp boundary condition for interpolation. Intermediate observer
heights are then interpolated as desired using splines.

The amplification after extinction for a given observatory may now be computed.
This is done by evaluating the \lowtran\ spectral interpolator at a depth equal
to the depth traveled by the lensed rays, which is itself a function of
wavelength. Since \lowtran\ assumes ground-to-space (although ``ground'' here
is really just a user-chosen altitude), then space-to-ground-to-space
transmission will simply be the self-product. Finally, this function is
then multiplied by the chromatic amplification function for the lunar
observatory.

\begin{figure}
\begin{center}
\includegraphics[width=1.05\columnwidth,angle=0,clip=true]{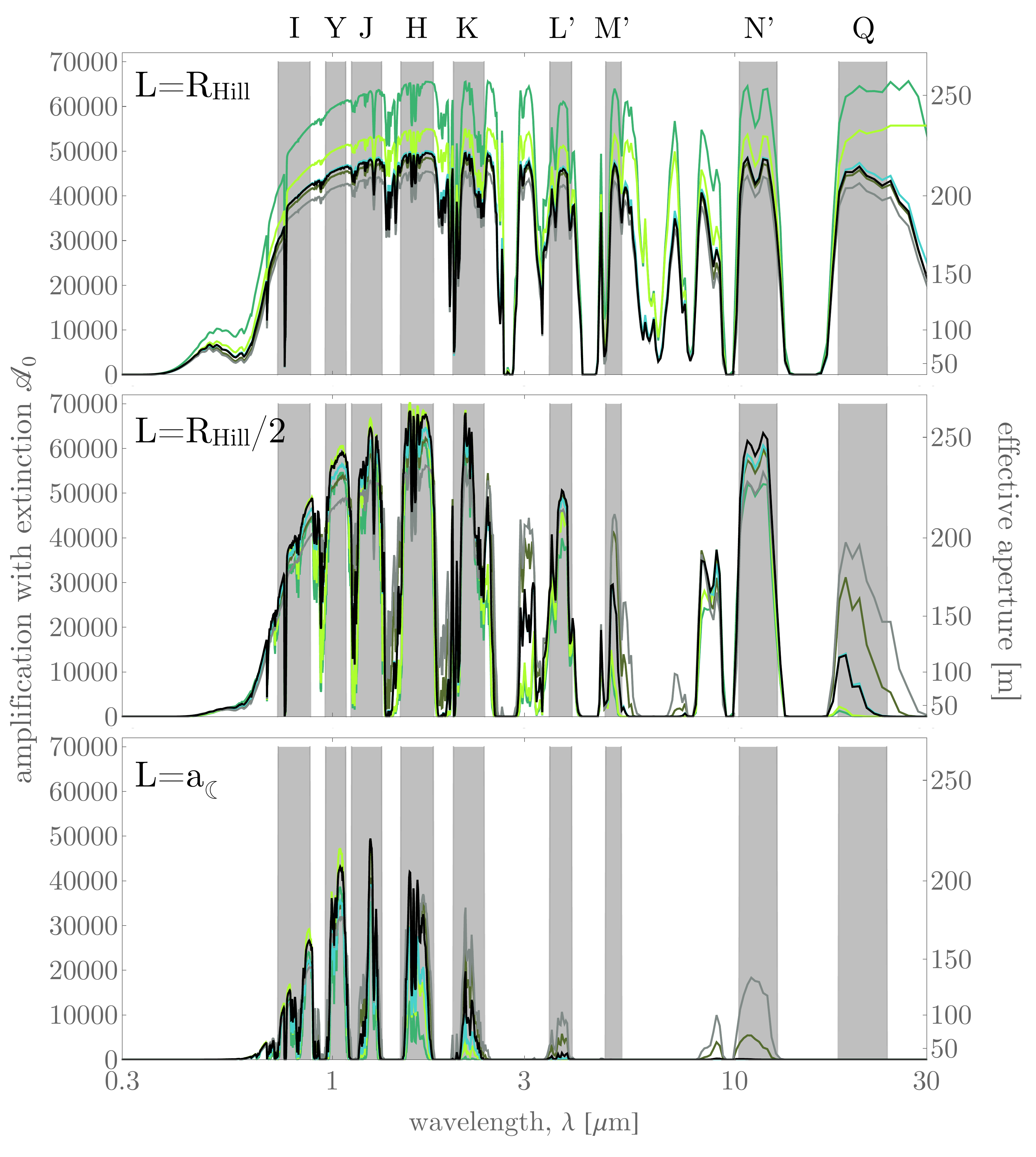}
\caption{
Amplification after extinction expected for a 1\,metre diameter telescope at
the Earth's Hill radius (top), half the Hill radius (middle)
and the Moon's separation (bottom). Six atmosphere models are shown
(same color coding as Figure~\ref{fig:focus}), which control
temperature-pressure profiles (and thus refractivity profile) as well as the
extinction computed using \lowtran. All models assume no clouds. Standard
photometric filters highlighted in gray, except for L, M and N which are
slightly offset to encompass the optimal regions.
}
\label{fig:extinction}
\end{center}
\end{figure}

As a test of the \lowtran\ model, the transmission was converted to
an equivalent atmospheric extinction coefficient for some common optical
filters. The extinction coefficient for B-band was found to be 0.45,
V-band 0.28, R-band 0.19, I-band 0.086 and H-band 0.080. These all
line-up with typical coefficients for a good observing site\footnote{
See \thisurl\ for a pedagogical description of extinction coefficients
and some typical values.
} 

Figure~\ref{fig:extinction} shows the amplification for 1\,metre terrascope
observatory after accounting for the \lowtran\ extinction. Despite the
extinction, amplification up to 70,000 remains feasible. One can
see from Figure~\ref{fig:extinction} that extinction is severe for detectors
at the Moon's orbital radius, since lensed rays need to travel deep through
the Earth's atmosphere - just a couple of km (see Figure~\ref{fig:ampplot}).
As we move out in orbital radius, sufficient lensing is obtained at higher
altitudes thereby reducing the effect of atmosphere extinction, with
clear benefits to such detectors.

\subsection{Interception by clouds}
\label{sub:onaxisclouds}

The grazing nature of the terrascope lensed rays means that interception
by clouds has the potential to dramatically attenuate the overall
transmission through the atmosphere. Even wispy high-altitude cirrus clouds,
with an optical depth of ${\sim}0.1$ and \SI{1}{\kilo\metre} thickness scale,
can appear completely opaque to terrascope rays since the path length can be
up to ${\sim}$\SI{100}{\kilo\metre}. This simplifies the analysis, since one
can simply assume that encountering any kind of cloud leads to zero
transmission. The real question is then what is the frequency which rays
intercept a cloud?

It is important to recall that for rays lensed onto a Hill sphere
terrascope, the deepest altitude penetrated by the ray is
\SI{13.7}{\kilo\metre} (see Figure~\ref{fig:lunar}), and at this altitude
there are almost no clouds. Thus, if $L \sim F$, then the $(D-R)\sim0$ and lensed
rays will have to traverse not only a large airmass but also most likely
intercept opaque clouds during their journey. On the other hand, observatories
away from $F$ require less deflection and thus need not travel so deep through
the Earth's atmosphere, largely avoiding clouds.

The relationship between $L$ and $(D-R)$ is well-constrained from our
simulations. The first thing to highlight is that redder than about a micron,
the refraction is almost achromatic and thus the lensing depth is approximately
constant for a given $L$ (this is apparent from Figure~\ref{fig:lunar}). Thus,
one can simply take $\lim_{\lambda \to \infty} (D-R)$ as an excellent
approximation for wavelengths redder than a micron. The second thing to
highlight is that if one varies $L$ from $F$ out to $R_{\mathrm{Hill}}$ in 100
uniform steps, the relationship is tight and monotonic, empirically found to be
described by

\begin{align}
\lim_{\lambda \to \infty} (D-R) \simeq a_0 (1 - a_1 e^{-L/a_2}),
\label{eqn:empiricaldepth}
\end{align}

where for the US Standard Atmosphere 1976 model one obtains
$a_0=$\SI{15.54}{\kilo\metre} $a_1 = 1.829$ and
$a_2 = $\SI{551,100}{\kilo\metre} (to four significant figures). 

To estimate the effect of clouds, this work uses data from the High-resolution
Infrared Radiation Sounder (HIRS) satellite instrument. Statistical properties
of clouds have been catalogued with multi-year observations taken from polar
orbit and are have been described in the literature \citep{wylie:1994,wylie:1998}.
This work uses the data made available at \hirs. Within a field of view of approximately
\SI{20}{\kilo\metre} by \SI{20}{\kilo\metre}, HIRS determines the effective
cloud fraction, $N \epsilon$, where $N$ is the frequency of clouds and
$\epsilon$ is the emissivity, which approximately equals one minus the
transmission, $T$.

\begin{figure*}
\begin{center}
\includegraphics[width=17.0cm,angle=0,clip=true]{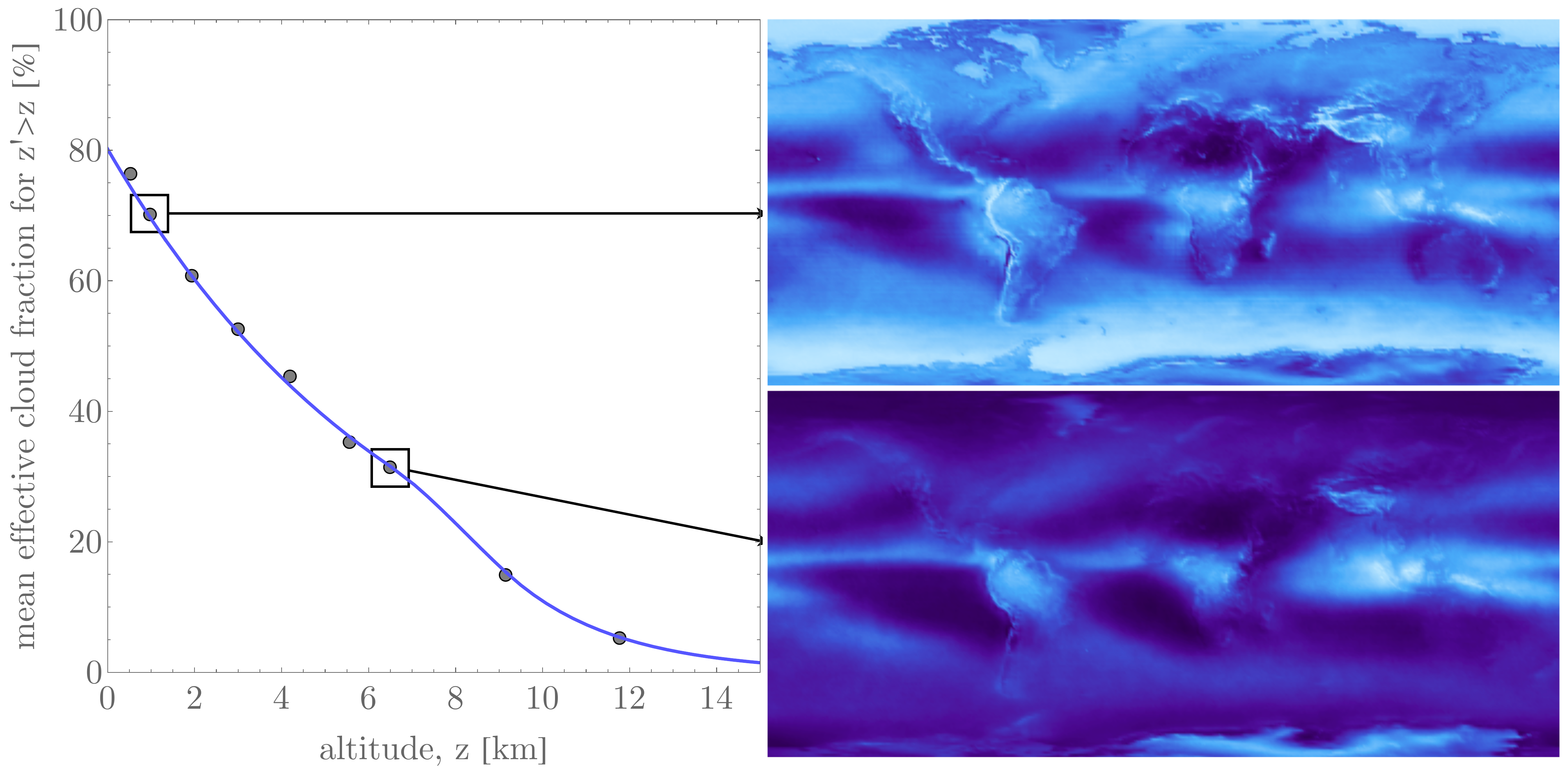}
\caption{
Left: Effective cloud fraction, $N \epsilon$, averaged over all months
and locations as measured over 11-years by HIRS, as a function of altitude
\citep{wylie:1994,wylie:1998}. Right: Two example cloud maps from the
data plotted on the left. Note how high altitude clouds are much more common
around the equatorial regions.
}
\label{fig:cloudmap}
\end{center}
\end{figure*}

Averaging over all longitudes, latitudes and months, the average effective
cloud fraction for all clouds below a pressure level of \SI{950}{\milli\bar} is
76.6\%. At or below a pressure level of \SI{200}{\milli\bar} the effective
cloud fraction has dropped to 5.4\%. The global averages are shown in
Figure~\ref{fig:cloudmap} where pressure levels have been converted to
altitudes. It is found that the nine available data points, for any given
location, are well described by a broken power-law, with a break at around one
scale height (as shown by the smooth function overplotted in the left panel of
Figure~\ref{fig:cloudmap}).

To generalize the HIRS data to arbitrary locations and altitudes, the
data set is first interpolated to a regularized grid then each location
fitted using the broken power-law. The interpolation is necessary because
no data is available north of \SI{84}{\degree} latitude or south of 
\SI{-84}{\degree}. To interpolate, longitudinal great circles are drawn
around the Earth in one-degree intervals and then the data is wrapped
around to ensure a continuous periodic function. A Gaussian Process with
a Mattern-3/2 kernel is trained at each pressure level across all
latitudes and used to fill in the missing latitudes. The broken power-law
is then fitted to each one-square degree location independently, where
the free parameters are two slopes, one offset and one transition point.

To simplify the analysis, only $L>R_{\mathrm{Hill}}/2$
is considered in what follows, meaning that $\lim_{\lambda \to \infty}
(D-R)>$\SI{8.2}{\kilo\metre}. At these altitudes, only high-altitude cirrus
clouds are present.

With these points established, it is now possible to estimate the impact
of clouds on terrascope rays. It is stressed that the following is
an approximate estimate and more detailed cloud modeling would be encouraged
in future work to refine the estimate made here. The purpose of this section
is to merely gauge the approximate feasibility of a terrascope when including
clouds.

If one assumes a terrascope detector orbiting in the Earth's equatorial plane,
then lensed rays will be described by a great circle of constant longitude
(or really one constant longitude plus another offset by \SI{180}{\degree}).
Since the HIRS public data used here has a resolution of \SI{1}{\degree}, one
can draw 180 such great circles - representing different rotational phases of
the Earth. Working in the equatorial plane is not only a simplifying assumption
but also minimizes the impact of high altitude clouds which are more frequent
at equatorial regions (see Figure~\ref{fig:cloudmap}).

For each great circle, there are 360 different locations (spread across
latitude) sampled in the (interpolated) HIRS data. Since a terrascope
detector located at $L=R_{\mathrm{Hill}}/2$ has focused red rays which 
traverse a depth of $(D-R) =$\SI{8.229}{\kilo\metre}, at each
location one can evaluate the cumulative effective cloud fraction above this
altitude using the broken power-law described earlier. Effective cloud fraction
is not equal to cloud frequency. Fortunately, for high altitude clouds
($>$\SI{6}{\kilo\metre}), the approximate relationship
$N \simeq \tfrac{1}{2} N \epsilon$ may be used \citep{wylie:1998}. Thus, at
each of the 360 points along the great circle, the cloud frequency for all
clouds above \SI{8.2}{\kilo\metre} altitude can be estimated.

Since the cumulative fraction is defined as all altitudes above altitude
$z$, and a terrascope ray indeed is forced to pass through all altitudes
above $z$, the cloud frequency $N$ may be interpreted as a time-averaged
transmission fraction for the depth $(D-R)$. The total transmission can
now be estimated by simply averaging over all such values along the
great circle.

\begin{figure}
\begin{center}
\includegraphics[width=1.05\columnwidth,angle=0,clip=true]{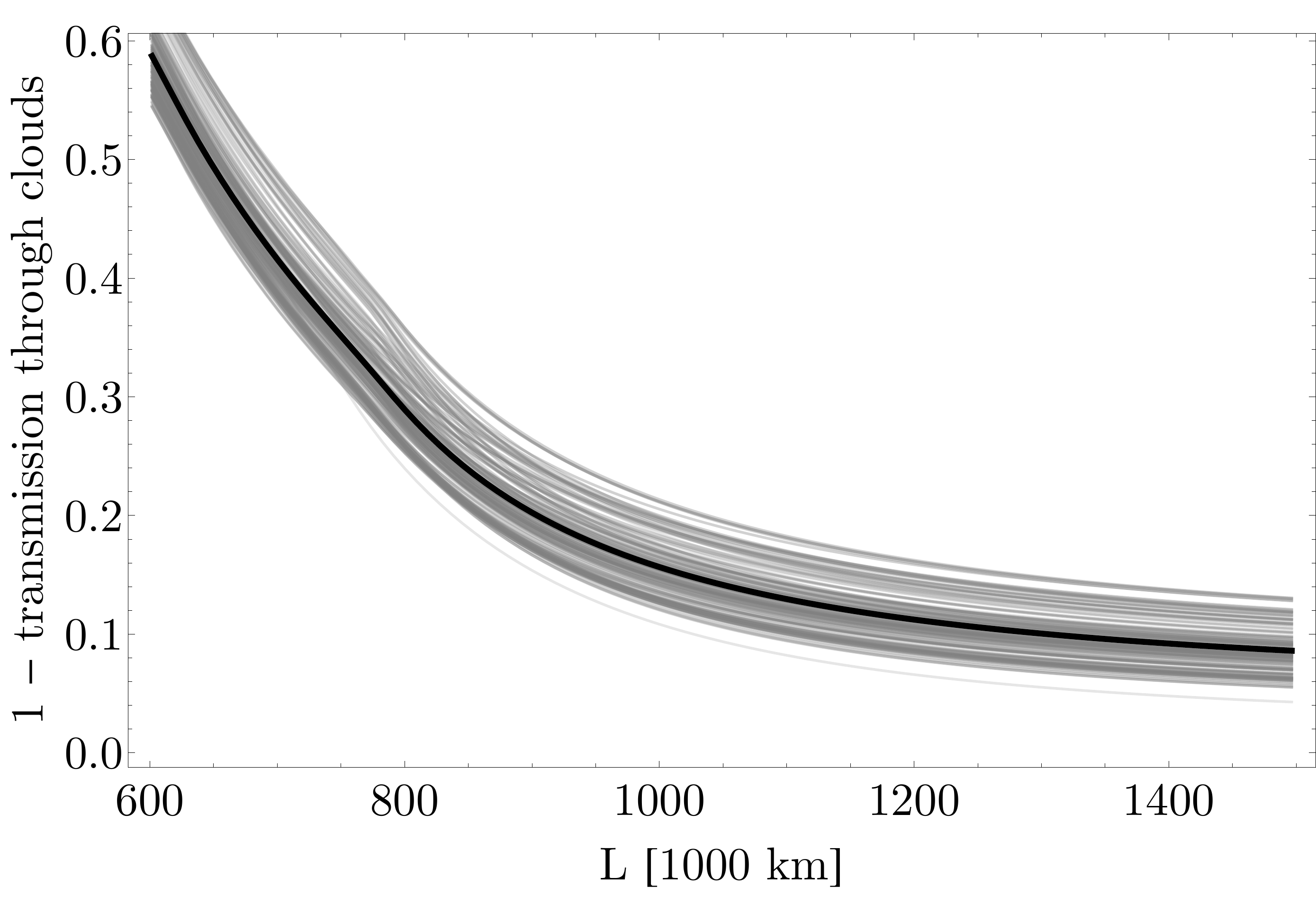}
\caption{
Estimated transmission through the Earth's atmosphere due to clouds for a
terrascope detector at a distance $L$. At one Hill radius
(\SI{1500000}{\kilo\metre}), lensed rays travel no deeper than
\SI{13.7}{\kilo\metre} and thus largely avoid clouds, thereby losing less
than 10\% of the lensed light.
}
\label{fig:cloudextinction}
\end{center}
\end{figure}

The results of this calculation are shown in Figure~\ref{fig:cloudextinction}.
A clear exponential trend is apparent in the results, highlighting as
expected how more distant terrascope observatories are less affected by
clouds. For the lowest $(D-R)$ allowed by our model, of \SI{6}{\kilo\metre},
$L=$\SI{600,000}{\kilo\metre} and the average cloud transmission is 41.4\%.
Moving out to $R_{\mathrm{Hill}}/2$, the situation is decidedly better with an
average transmission of 64.9\% and by the time $L=R_{\mathrm{Hill}}$, 91.9\% of
the lensed rays make it through the atmosphere unimpeded by clouds. In
conjunction with the earlier extinction calculations, these
results strongly suggest that a terrascope detector as close to
$L=R_{\mathrm{Hill}}$ as possible would optimize the setup.

\subsection{Off-axis lensing}
\label{sub:offaxisresults}

\begin{figure*}
\begin{center}
\includegraphics[width=17.0cm,angle=0,clip=true]{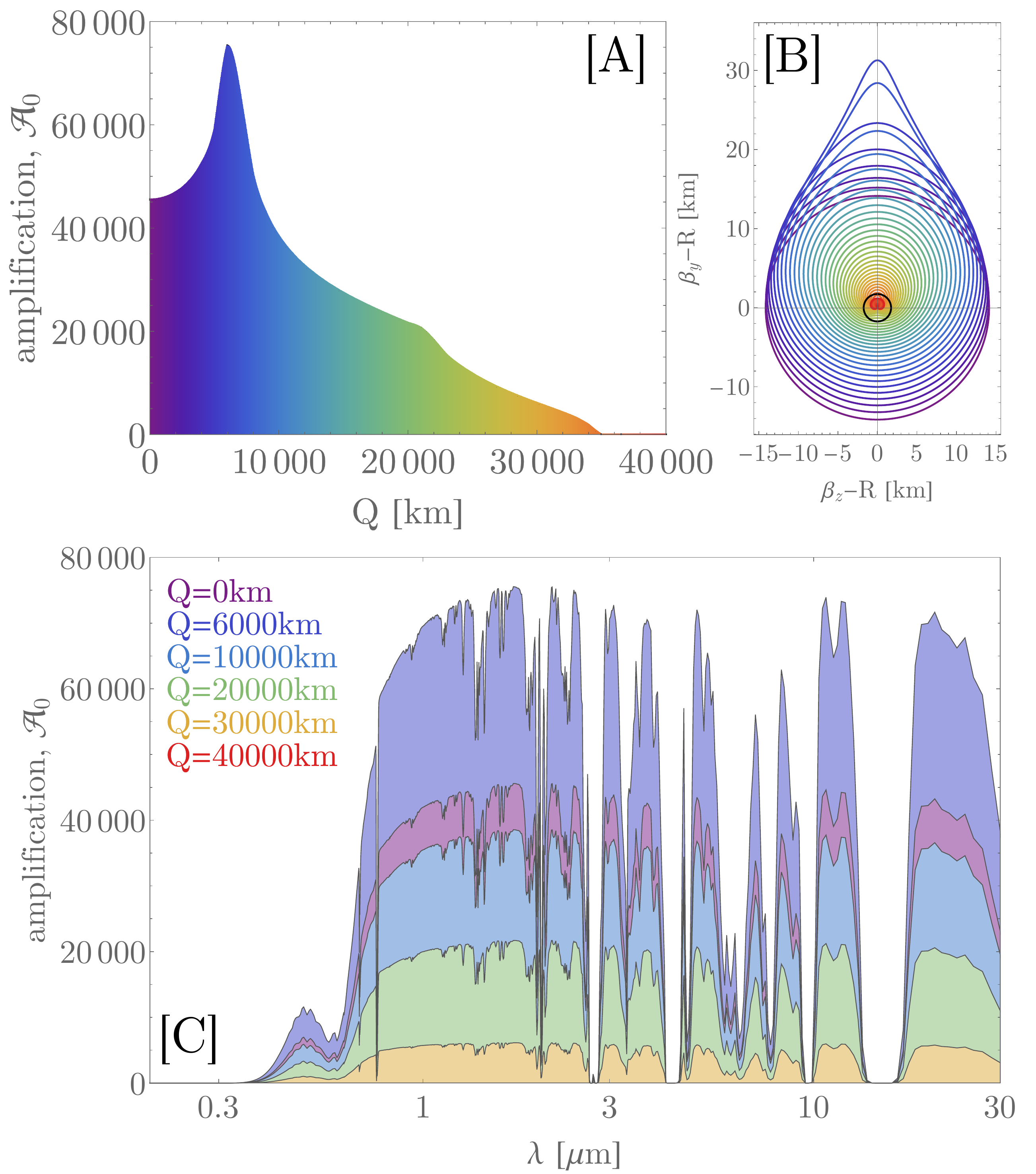}
\caption{
Off-axis lensing through the terrascope. Panel [A] shows the amplification
after extinction for $\lambda=$\SI{1.74}{\micro\metre}. Panel [B] shows
the simulated lensed images for 41 evenly spaced offset distances from
$Q=$\SI{0}{\kilo\metre} to $Q=$\SI{40000}{\kilo\metre}. Panel [C] shows the
spectral amplification at six different offset distances.
}
\label{fig:offaxisresults}
\end{center}
\end{figure*}

Off-axis lensing was calculated using the method described in
Section~\ref{sub:offaxis}. The terrascope detector is fixed to a distance
of $L=R_{\mathrm{Hill}}$ and to $W=$\SI{1}{\metre} in what follows. The shape
of the lensed source around the Earth was computed for the US Standard
Atmosphere 1976 model at various off-axis distances ranging from
$Q=$\SI{0}{\kilo\metre} to $Q=$\SI{40000}{\kilo\metre} to
\SI{1000}{\kilo\metre} steps. As with the images shown in
Figure~\ref{fig:shapes}, the rings are often egg-shaped (see
Figure~\ref{fig:offaxisresults}) and thus the area was calculated through
numerically integration along 2000 uniformly spaced choices of $\phi$, yielding
an amplification value. The calculation was then repeated across the same grid
of wavelengths used earlier in Section~\ref{sub:trainingset}.

The amplification computed above describes the idealized case with
no extinction. Since the shape is saved in each simulation, this information
can be used to estimate the fraction of lost light due to clear sky
extinction and also that of clouds, using the same methods described
earlier in Sections~\ref{sub:onaxisextinction} \& \ref{sub:onaxisclouds}.
Since the depth varies as a function of $\phi$ along the ring, the overall
transmission is given by the amplification multiplied by the mean of the
extinction over all 2000 phase points (where extinction here includes both
the clear sky and cloud components). The resulting amplifications from
this process are shown in Figure~\ref{fig:offaxisresults}.

For offsets of $Q>$\SI{18900}{\kilo\metre}, the amplification has dropped to
less than half that as on-axis. This may be converted into a timescale by
nothing that at one Hill radius, a satellite would have a tangential velocity
of ${\simeq}$\SI{0.5}{\kilo\meter/\second}. Accordingly, the lensing timescale
would be ${\sim}$\SI{20}{\hour} including both sides of the off-axis lensing,
or roughly a day. During this time, the target has moved by approximately
\SI{1.4}{\degree} on the sky.

\section{Discussion}
\label{sec:discussion}

\subsection{Magnification}
\label{sub:magnification}

The calculations described thus far concern the amplification in flux of a
distant source with a terrascope, but not the magnification in angular
size of a source. Variations in the Earth's atmosphere already
present a major limiting factor in the resolving power of ground-based
telescope and thus one should expect it to be an even greater for the terrascope.
Even a Hill sphere observatory, exploiting stratospheric lensing, will observe
rays that have traversed through ${\sim}$20 airmasses (see
Figure~\ref{fig:interp}) and thus seeing will be of order tens of arcseconds.

Consider a distant object whose light arrives at the Earth such that the
object subtends an angle $\phi$. This light is refracted and arrives at the terrascope
detector subtending an angle $\theta = \Delta + \phi$, where $\Delta$ is the
deflection angle. The magnification is $\theta/\phi = (\Delta/\theta - 1)^{-1}$.
Since $\theta \simeq b/L$ and $\Delta = \Delta_0 e^{-(b-R)/H_{\Delta}}$, then
magnification is approximately $[(\Delta_0 L/b) e^{-(b-R)/H_{\Delta}} - 1]^{-1}$.
Magnification thus theoretically tends to infinity, representing a caustic, as
$b$ approaches the value for maximum amplification. This caustic behavior
not surprisingly echoes the situation of gravitational lensing, although
practically realizing this magnification would not possible given the
atmospheric disturbance.

\subsection{Separation of nearby sources}
\label{sub:contamination}

The fact that an off-axis source still produces significant lensing as much
as \SI{1.4}{\degree} is useful, because lensing events occur over a prolonged
timescale, but also potentially problematic. This is because it indicates that
nearby sources, within a degree, will have some fraction of their light also
lensed onto the detector and thus might be concerned about blending.

Consider the on-axis lensing scenario but with an additional source
offset by an angle $\theta$ on the sky. The line connecting the contaminating
source and the detector does not pass through the Earth's center (as with
on-axis lensing) but rather is offset by a distance $Q$, meaning that
$\theta = Q/L$. The on-axis lensed sourced travels through the Earth's
atmosphere at an impact parameter of $b_{\mathrm{mid}}$, which is equivalent to
$\lim_{Q\to0} b_{\mathrm{mid}}(Q)$, whereas the offset source has
$b_{\mathrm{mid}}(Q)$. These two lensed images appear separated in the
atmosphere, as seen from the detector, by an angle of $\alpha$ given by

\begin{align}
\alpha &= \frac{ (\lim_{Q\to0} b_{\mathrm{mid}}(Q)) - b_{\mathrm{mid}}(Q) }{L}
\end{align}

Accordingly, the apparent angular separation decreases from $\theta$ to $\alpha$
by the ratio

\begin{align}
\frac{\alpha}{\theta} &= \frac{ (\lim_{Q\to0} b_{\mathrm{mid}}(Q)) - b_{\mathrm{mid}}(Q) }{Q}.
\end{align}

Using the numerical results from Section~\ref{sub:offaxisresults}, this ratio
can be computed for any given wavelength and at any given phase angle around
the Earth. For $Q>$\SI{20000}{\kilo\meter}, all phase angles converge to a
$\tfrac{\alpha}{\theta}$ ratio of one over a few thousand. Since
the detector has a diffraction limited angular resolution of
$1.22\tfrac{\lambda}{W}$, then the source separation ability of the terrascope
will be $\sim \tfrac{\lambda H_{\Delta}}{W R}$. This is 
$\simeq 0.25$\,milliarcseconds for a 1\,metre detector at
\SI{1}{\micro\metre} and thus is a factor of $1.22 H_{\Delta}/W$ improved.

Aside from resolving a contaminant through angular separation, it may be
possible to separate sources (to some degree) based on the distinct temporal
lensing light curves that emerge due to the differing geometries.

\subsection{Atmospheric radiance}
\label{sub:onaxisradiance}

The Earth's atmosphere is luminous from airglow, scattering and thermal
emission and this radiance poses an obstacle to the terrascope. By using a
shade adapted for the Earth, it may be possible to remove flux from the
Earth's disk, which greatly outshines the sky brightness. A simple shade
would need to be offset from the detector by a distance of $L [b/(W+b)]$
in order to occult the Earth's disk and have a radius of $R [W/(W+b)]$.
For all detectors with $W<$\SI{0.209}{\metre} (corresponding to an effective
telescope diameter of \SI{96.9}{\metre}), the effective collecting area of
the terrascope exceeds the size of the Earth-shade. It may be still economical
to go beyond \SI{0.2}{\metre} since the cost of a shade is expected to be
cheaper than a mirror.

Scattering from the upper atmosphere will be ever present and represents
a source of background (rather than necessarily a source of noise). This
background will be strongly dependent upon the relative position of the
Sun during the observations. Let us denote the angle subtended from the
Earth to the terrascope detector to the Sun as $\Theta$. If
\SI{0}{\degree}$<\Theta<$\SI{90}{\degree} or
\SI{270}{\degree}$<\Theta<$\SI{360}{\degree}, then the Sun will appear directly
in view to the detector excluding observations during this time. If
\SI{90}{\degree}+\SI{18}{\degree}$<\Theta<$\SI{270}{\degree}-\SI{18}{\degree},
then one side of the Earth will be in astronomical twilight where scattered
sunlight cannot interfere (except at the instant of $\Theta=\pi$). If the
observatory is exactly in the ecliptic plane, then at any one time during this
range in $\Theta$ exactly one half of the Earth's circumference will be in
astronomical twilight.

Accordingly, it is estimated here that the actual amplification from a
terrascope will be one half of that depicted in the various figures
throughout. This assumes that any part of the Earth which is illuminated
will have a background component that is simply not removable. However,
more detailed calculations than possible here may be able demonstrate that
at least some fraction of this lost capability can be recovered through
background suppression strategies, such as leveraging polarization,
wavelength information, and temporal light curve variations. These
are undoubtedly technical challenges for a realized terrascope but effort
should be encouraged to explore overcoming them given the very large
gains potentially given by such a system.

\subsection{Atmospheric stability}

The refractivity of air at a specific altitude will vary as a function of
position and time in a realistic atmosphere. It is argued here that so
long as the terrascope detector is a significant distance away from the
inner focal point, these variations will not affect the amplification
factor in a meaningful way. Consider a particular location where there
is a increase in pressure at altitude $z$ compared to the typical pressure
at altitude $z$. This causes the refractivity to increase and thus light
traveling at that location will now refract too much and miss the detector
at distance $L$. However, there must be an altitude $z'>z$ where the pressure
decreases back down to the typical pressure, thereby refracting light back
onto the detector. In this way, the perfect circular ring image is distorted
into an irregular ring - but the thickness of the ring is the same and
thus the amplification is unchanged.

\subsection{Pointing}

Since an off-axis source still causes significant lensing at \SI{1.4}{\degree}
for a Hill sphere terrascope, this denotes the approximate angular band
on the sky suitable for observation. This represents just under one
percent of the sky. The orbital plane of the detector is a free parameter
but ecliptic observing minimizes the affect of high altitude clouds and Solar
scattering, as well as providing the densest field of targets. Pointing is
naturally limited to whatever happens to be behind the Earth at any given time,
although fleets of terrascope detectors could increase the coverage as needed.

\subsection{Radio terrascope}

The calculations of extinction in this work strictly assume optical/infrared
light. Moving further out into the radio offers two major advantages though.
First, extinction due to clouds can be largely ignored, allowing for detectors
much closer including on the lunar surface. Second, Solar scattering is far
less problematic in the radio and indeed it is typical for radio telescopes to
operate during daylight phases. The simple refraction model of this work was
extended to the radio and indeed the amplification was estimated to be largely
achromatic beyond a micron. Nevertheless, the model did not correctly account
for the radio refractivity as a function of humidity, nor the impact of the
ionosphere on lensed rays. Accordingly, a radio terrascope may be an excellent
topic for further investigation. It should be noted though that a disadvantage
of a giant radio receiver in space is that humanity already regularly builds
large receivers on Earth at much lower expense than their optical counterparts.
Thus, the benefit of going into space for radio observations may not prove
ultimately economical.

\acknowledgments

DMK is supported by the Alfred P. Sloan Foundation. Thanks to members of the
Cool Worlds Lab and the NASA Goddard Institute for Space Science group for
useful discussions in preparing this manuscript. Thanks to Jules Halpern,
Duncan Forgan, Caleb Scharf and Claudio Maccone for reviewing early
drafts and discussions of this work. Special thanks to Tiffany Jansen for
her assistance with coding questions. Finally, thank-you to the anonymous
reviewer for their constructive feedback.

%

\vspace{5mm}


\software{\lowtran\ \citep{lowtran:1988}
          }






\begin{thebibliography}{99}
\bibitem[\protect\citeauthoryear{Birch \& Downs}{1994}]{birch:1994} 
Birch, K.~P. \& Downs, M.~J., 1994, Metrologia, 31, 315.
\bibitem[\protect\citeauthoryear{Cassini}{1740}]{cassini:1740} 
Cassini, J.~D., 1740, ``Tables astronomiques'', p. 34 
\bibitem[\protect\citeauthoryear{Conrad et al.}{1969}]{apollo12:1969} 
Conrad, C., Gordon, R.~F., Bean. A.~L., 1969,
``Earth Eclipses the Sun - Apollo 12'', NASA JPL https://moon.nasa.gov/resources/199/earth-eclipses-the-sun-apollo-12/
\bibitem[\protect\citeauthoryear{Eddington}{1919}]{eddington:1919} 
Eddington A.~S., 1919, Obs, 42, 119.
\bibitem[\protect\citeauthoryear{Edl\'en}{1966}]{edlen:1966} 
Edl\'en, B., 1994, Metrologia, 2, 71.
\bibitem[\protect\citeauthoryear{Einstein}{1916}]{einstein:1916} 
Einstein, A., 1916, Annalen der Physik, 49, 769.
\bibitem[\protect\citeauthoryear{Elvis}{2016}]{elvis:2016} 
Elvis, M., 2016, ``The Crisis in Space Astrophysics and Planetary
Science: How Commercial Space and Program Design Principles will let us
Escape'', Frontier Research in Astrophysics II (arXix e-prints:1609.09428).
\bibitem[\protect\citeauthoryear{Heidmann \& Maccone}{1994}]{heidmann:1994} 
Heidmann, J. \& Maccone, C., 1994, Acta Astron., 32, 409.
\bibitem[\protect\citeauthoryear{Hubbard et al.}{1987}]{hubbard:1987} 
Hubbard, W.~B., Nicholson, P.~D., Lellough, E., et al., 1978, Icarus, 72, 635.
\bibitem[\protect\citeauthoryear{Kneizys et al.}{1988}]{lowtran:1988} 
Kneizys, F.~X., Shettle, E.~P., Abreu, L.~W., Chetwynd, J.~H., Anderson, G.~P.,
Gallery, W.~O., Selby, J.~E.~A., Clough, S.~A., 1988, ``User's guide to LOWTRAN7'',
Air Force Geophysics Lab, Tech. Rep.
\bibitem[\protect\citeauthoryear{Kraus}{1986}]{kraus:1986} 
Kraus, J.~D., 1986, Radio Astronomy, Cygnus-Quasar Books, Powell, Ohio, p. 6.
\bibitem[\protect\citeauthoryear{Monnier}{2003}]{monnier:2003} 
Monnier, J.~D., 2003, Reports on Progress in Physics, 66, 789.
\bibitem[\protect\citeauthoryear{National Geophysical Data Center}{1992}]{US:1976} 
National Geophysical Data Center: U.S. standard atmosphere (1976), 1992, Planet. Space Sci., 40, 553.
\bibitem[\protect\citeauthoryear{Rasmussen \& Williams}{2006}]{gp2} 
Rasmussen, C.~E. \& Williams, C., 2006, ``Gaussian Processes for Machine Learning'',
MIT Press, Cambridge
\bibitem[\protect\citeauthoryear{Stein}{1999}]{gp1} 
Michael, S.~L., 1999, ``Statistical Interpolation of Spatial Data: Some Theory for Kriging'',
Springer, New York
\bibitem[\protect\citeauthoryear{Turyshev \& Andersson}{2003}]{turyshev:2003} 
Turyshev, S.~G. \& Andersson, B.-G., 2003, MNRAS, 341, 577.
\bibitem[\protect\citeauthoryear{van Belle et al.}{2004}]{belle:2004} 
van Belle, G.~T., Meinel, A.~B. \& Meinel, M.~P., 2004, in SPIE Conf.
Ser. 5489, ed. J. M. Oschmann, Jr., 563.
\bibitem[\protect\citeauthoryear{von Eshleman}{1979}]{von:1979} 
von Eshleman R., 1979, Sci, 205, 1133.
\bibitem[\protect\citeauthoryear{Wang}{1998}]{wang:1998} 
Wang, Y., 1998, Proc. SPIE, 3356, 665
\bibitem[\protect\citeauthoryear{Wylie et al.}{1994}]{wylie:1994} 
Wylie, D.~P., Menzel, P.~W., Woolf, H., M. \& Strabala, K.~I., 1994,
Journal of Climate, 31, 1972.
\bibitem[\protect\citeauthoryear{Wylie \& Menzel}{1998}]{wylie:1998} 
Wylie, D.~P. \& Menzel, P.~W., K.~I., 1994, Journal of Climate, 12, 170.
\end{thebibliography}
\end{document}